\documentclass[letterpaper,12pt,oneside,onecolumn]{article}

\usepackage{geometry}
\usepackage{titlesec}
\titleformat{\section}[hang]{\normalfont\normalsize\bfseries}{\thesection}{12pt}{\centering}
\titleformat{\subsection}[display]{\normalfont\normalsize}{\thesubsection}{12pt}{\underline}
\titleformat{\subsubsection}[runin]{\normalfont\normalsize}{\thesubsubsection}{12pt}{\underline}

\newcommand{\TPCI}[3]{
\begin{flushright}
    \begin{scriptsize}
        \textbf{#1} \textit{#2} \\
        \textbf{\textit{TMS (The Minerals, Metals \& Materials Society), #3}}\\
    \end{scriptsize}
\end{flushright}
}

\newcommand{\PaperTitle}[1]{
\begin{center}
    \begin{large}
        \textbf {#1} \\
    \end{large}
\end{center}
}

\newcommand{\AuthorList}[1]{
\begin{center}
    {#1} \\
\end{center}
}

\newcommand{\AuthorAffiliations}[1]{
\begin{center}
    {#1} \\
\end{center}
}

\usepackage[latin1]{inputenc}
\usepackage[sort]{cite}
\usepackage{amsmath}
\usepackage[]{amssymb}
\usepackage{xspace}
\usepackage{graphicx}
\usepackage{subfigure}
\newcommand{\nSc}{n_{\mathrm{Sc}}}
\newcommand{\nZr}{n_{\mathrm{Zr}}}
\newcommand{\nX}{n_{\mathrm{X}}}
\newcommand{\ud}{\mathrm{d}}

\newcommand{\etal}{\textit{et al}.\@\xspace}
\newcommand{\ie}{\textit{i.e.}\@\xspace}
\newcommand{\abinitio}{{ab initio \xspace}}

\newcommand{\me}{\mathrm{e}}
\DeclareMathOperator{\q}{q}
\newcommand{\xeq}{x_{\mathrm{X}}^{eq}}
\newcommand{\xO}{x_{\mathrm{X}}^{0}}
\newcommand{\oun}{\omega^{(1)}_{\mathrm{AlX}}}
\newcommand{\odeux}{\omega^{(2)}_{\mathrm{AlX}}}
\newcommand{\muAl}{\mu_{\mathrm{Al}}}
\newcommand{\muZr}{\mu_{\mathrm{Zr}}}
\newcommand{\muSc}{\mu_{\mathrm{Sc}}}

\begin{document}

\TPCI{Solid-Solid Phase Transformations in Inorganic Materials}{Edited by}{2005}

\PaperTitle{Precipitation in Al-Zr-Sc alloys: \\
a comparison between kinetic Monte Carlo, cluster dynamics \\
and classical nucleation theory}

\AuthorList{Emmanuel Clouet,$^1$ 
Maylise Nastar,$^1$ 
Alain Barbu,$^1$ Christophe Sigli,$^2$ Georges Martin$^3$}

\AuthorAffiliations{$^1$Service de Recherches de Métallurgie Physique, CEA/Saclay\\
91191~Gif-sur-Yvette, France

$^2$Alcan CRV, B.P.~27, 38341~Voreppe, France

$^3$Cabinet du Haut-Commissaire, CEA/Siège, 31-33 rue de la Fédération \\
75752~Paris~cedex 15, France}

%\Keywords{Precipitation, kinetic Monte Carlo, cluster dynamics, nucleation.}

\section{Abstract} 
Zr and Sc precipitate in aluminum alloys to form the Al$_3$Zr$_x$Sc$_{1-x}$ compound
which, for low supersaturations of the solid solution, exhibits the L1$_2$ structure. 
The aim of the present study is to model at an atomic scale the kinetics of precipitation
and to build mesoscopic models so as to extend the range of supersaturations 
and annealing times that can be simulated up to values of practical interest.
In this purpose, we use some \abinitio calculations and experimental data
to fit an Ising type model describing thermodynamics of the Al-Zr-Sc system.
Kinetics of precipitation are studied with
a kinetic Monte Carlo algorithm based on an atom-vacancy exchange mechanism.
Cluster dynamics is then used to model at a mesoscopic scale all the different stages
of homogeneous precipitation in the two binary Al-Zr and Al-Sc alloys. 
This technique correctly manages to reproduce both the kinetics of precipitation
simulated with kinetic Monte Carlo as well as experimental observations.
Focusing on the nucleation stage, it is shown that classical theory well applies
as long as the short range order tendency of the system is considered. 
This allows us to propose an extension of classical nucleation theory
for the ternary Al-Zr-Sc alloy.

\section{Introduction} 

Transition elements like Mn, Cr, Zr or Sc are usually added  to aluminum alloys
so as to form small ordered precipitates which increase the tensile strength
and inhibit recrystallization.
As these properties directly depend on the density and the size of the particles,
an accurate control of the precipitation process is necessary 
so as to optimize the material. 
In particular, one can use a combined addition of solute elements so as to change
the precipitate distribution and thus improve the alloy properties.
Such an example for aluminum alloy is the addition of Zr and Sc elements.
Both elements, when added separately, increase the tensile strength 
and the recrystallization resistance, but the effect is even better 
for the combined addition as there are more precipitates which are smaller 
and less sensitive to coarsening \cite{YEL85,DAV96,TOR98,FUL03,RID04,ROY05b}.
So as to better understand the precipitation kinetics in Al-Zr-Sc 
we develop a multiscale approach combining atomic and mesoscopic simulations.

At the atomic scale, kinetic Monte Carlo (KMC) simulations are the suitable tool 
to study precipitation kinetics. 
Thanks to a precise description of the physical phenomenon leading to the evolution 
of the alloy, such simulations allow to predict the kinetic pathways in full details.
These predictions are valuable for a ternary alloy like Al-Zr-Sc 
because few information on the precipitation kinetics is available
whereas the diversity of the possible kinetic pathways is richer than for a binary alloy.
Nevertheless, one drawback of this atomic approach is the needed computational time:
only the early stages of precipitation and the high supersaturations can be simulated.
There is thus a need to develop mesoscopic models so as to study 
alloys and heat treatments compatible with technological requirements
and to allow a comparison between simulated kinetics and experimental data.
These mesoscopic models, which include cluster dynamics (CD) and the ones based
on classical nucleation theory (CNT), 
describe the alloy by the means of the cluster size distribution. 
Their input parameters are not always directly available
but can be obtained from the atomic model. 
Their main drawback is that they are not as predictive as atomic models,
especially for ternary alloys where one has to know the kinetic pathway
before modeling the precipitation kinetics.
Therefore, a multiscale approach combining atomic and mesoscopic simulations
sounds valuable. It allows to use KMC simulations 
to obtain the missing information required by mesoscopic models
and thus extend the range of supersaturations and heat treatments
that can be studied.

\section{Kinetic Monte Carlo (KMC)}
\subsection{Atomic model}

\subsubsection{Al-Zr-Sc thermodynamics.}
The system we want to model is the ternary Al-Zr-Sc alloy.
The aluminum solid solution has a face-centered-cubic (fcc) structure and 
the forming precipitates are Al$_3$Zr, Al$_3$Sc, and Al$_3$Zr$_x$Sc$_{1-x}$,
all precipitates having the L1$_2$ structure.\footnote{Al$_3$Zr and Al$_3$Zr$_x$Sc$_{1-x}$ 
for $0.8\lesssim x \leq1$ have the stable DO$_{23}$ structure 
\cite{PEARSON,YEL85,HAR02}. Nevertheless for small supersaturations of the solid solution,
precipitates with the L1$_2$ structure nucleate and grow first. Precipitates with
the DO$_{23}$ structure only appear for prolonged heat treatment and high enough
supersaturations \cite{ROB01,RYU69,NES72}.} 
This structure relies on an fcc underlying lattice. 
Therefore a rigid lattice model can be used so as to describe the different 
configurations of the system. 
This model is well suited to the Al-Zr-Sc alloy because the lattice parameters
of the forming precipitates are really close to that of the aluminum matrix
and precipitates remain coherent at least for diameters smaller than 10~nm. 
Therefore elastic relaxation as well as loss of coherency can be neglected 
in a study focusing on the first stages of the precipitation kinetics.

Atoms are thus constrained to lie on an fcc lattice configurations of which
are described by the occupation numbers $p^i_n$.
$p^{i}_{n}=1$ if the site $n$ is occupied by an atom of type $i$
and $p^{i}_{n}=0$ if not. Energies of such configurations are
given by an Ising model with first and second nearest neighbor
interactions. Thus, in our model, the energy per site of
a given configuration is
\begin{equation}
E = \frac{1}{2 N_s} \sum_{\substack{ n,m \\ i,j}}
\varepsilon_{ij}^{(1)} p^{i}_{n} p^{j}_{m} + \frac{1}{2 N_s}
\sum_{\substack{ r,s \\ i,j}} \varepsilon_{ij}^{(2)} p^{i}_{r}
p^{j}_{s} , \label{eqIsing}
\end{equation}
where the first and second sums respectively runs on all first and
second nearest neighbor pairs of sites, $N_s$ is the number of
lattice sites, $\varepsilon_{ij}^{(1)}$ and $\varepsilon_{ij}^{(2)}$ are
the respective effective energies of a first and second nearest
neighbor pair in the configuration \{$i,j$\}.

Such a rigid lattice model has been previously developed for 
Al-Zr and Al-Sc binary alloys \cite{CLO04}. 
First and second nearest neighbor interactions have been fitted
so as to reproduce the free energies of formation 
of Al$_3$Zr and Al$_3$Sc compounds in the L1$_2$ structure 
as well as Zr and Sc solubility limits in aluminum.\footnote{For
Al-Zr interaction, as we want to model precipitation of the metastable
L1$_2$ structure, we use the metastable solubility limit that 
we previously obtained from \abinitio calculations \cite{CLO02}.}
So as to study precipitation in the ternary Al-Zr-Sc system, 
this atomic model needs to be generalized by including interactions
between Zr and Sc atoms. As no experimental information is available,
we use \abinitio calculations to deduce the corresponding parameters.
The formation energies of 19 ordered compounds in the Al-Zr-Sc system
are calculated with the full potential linear muffin tin orbital method\cite{MET93}
in the generalized gradient approximation\cite{PER96}. 
We then use the inverse Connoly-Williams method\cite{CON83}
to obtain the two missing parameters of the atomic model, \ie 
$\varepsilon^{(1)}_{\mathrm{ZrSc}}$ and $\varepsilon^{(2)}_{\mathrm{ZrSc}}$
are fitted on this database of formation energies.
We obtain an order energy $\omega^{(1)}_{\mathrm{ZrSc}} 
= \epsilon_{\mathrm{ZrSc}}^{(1)} - \frac{1}{2}\epsilon_{\mathrm{ZrZr}}^{(1)}
-\frac{1}{2}\epsilon_{\mathrm{ScSc}}^{(1)} = 237$~meV corresponding to 
a strong repulsion when Zr and Sc atoms are first nearest neighbors
and a slight attraction when they are second nearest neighbors 
($\omega^{(2)}_{\mathrm{ZrSc}}=-2.77$~meV).
With such interactions, an ordered ternary compound Al$_6$ZrSc is stable 
at 0K. It is based on the L1$_2$ structure with Zr and Sc atoms being ordered
on the minority sublattice so as to form only attractive second nearest 
neighbor ZrSc interactions.
When increasing the temperature, this structure partially disorders
and leads to an Al$_3$Zr$_x$Sc$_{1-x}$ compound where $0 \leq x \leq 1$
is a variable quantity. This compound has a L1$_2$ structure too 
and the atoms on the minority sublattice can equally be Zr or Sc.
The order-disorder transition temperature estimated within Bragg-Williams
approximation is $T\sim70$K. Thus, at the temperatures we are interested in,
\ie above the ambient temperature, our atomic model predicts that only 
the Al$_3$Zr$_x$Sc$_{1-x}$ compound is stable. This agrees with experimental
observations \cite{YEL85,TOR90,HAR02} performed with transmission electron microscopy (TEM)
which show that precipitates in the ternary system have the structure 
described above.

\subsubsection{Al-Zr-Sc kinetics.}

We introduce in the Ising model atom-vacancy interactions for first nearest neighbors:
$\varepsilon^{(1)}_{\mathrm{AlV}}$, $\varepsilon^{(1)}_{\mathrm{ZrV}}$,
and $\varepsilon^{(1)}_{\mathrm{ScV}}$ are deduced from vacancy 
formation energy and binding energies with solute atoms in aluminum \cite{CLO04}.

Diffusion can then be described through vacancy jumps.
The vacancy exchange frequency with one of its twelve first nearest neighbors
of type $\alpha$ is given by
\begin{equation}
\Gamma_{\alpha\mathrm{-V}}=\nu_{\alpha}
\exp{\left(-\frac{E^{act}_{\alpha}}{kT} \right)} , 
\label{eq:rate}
\end{equation}
where $\nu_{\alpha}$ is an attempt frequency and the activation
energy $E^{act}_{\alpha}$ is the energy change required to move
the $\alpha$ atom from its initial stable position to the saddle
point position. It is computed as the difference between the
contribution $e^{sp}_{\alpha}$ of the jumping atom to the saddle
point energy and the contribution of the vacancy and of the
jumping atom to the initial energy of the stable position. This
last term is obtained by considering all bonds which are
broken during the jump.
The six kinetic parameters $\nu_{\alpha}$ and $e^{sp}_{\alpha}$
are fitted so as to reproduce Al self diffusion coefficient
and Zr and Sc impurity diffusion coefficients\cite{CLO04}.

\subsection{Atomic simulations}

\subsubsection{Kinetic Monte Carlo algorithm.}
We use time residence algorithm to run KMC
simulations. The simulation boxes contain $N_s=100^3$ or $200^3$
lattice sites and a vacancy occupies one of these sites. At each
step, the vacancy can exchange with one of its twelve first nearest
neighbors, the probability of each jump being given by
Eq.~(\ref{eq:rate}). The time increment corresponding to this
event is
\begin{equation}
\Delta t = \frac{1}{N_s(1-13x^0_{\mathrm{X}})
C_{\mathrm{V}}(\mathrm{Al}) }
\frac{1}{\sum_{\alpha=1}^{12}{\Gamma_{\alpha\mathrm{-V}}}},
\label{eq:time_inc}
\end{equation}
where $x^0_{\mathrm{X}}=x^0_{\mathrm{Zr}}+x^0_{\mathrm{Sc}}$
is the solute nominal concentration of the simulation box
and $C_{\mathrm{V}}(\mathrm{Al})$ the real vacancy concentration in
pure Al as deduced from energy parameters.

Criteria used to discriminate atoms belonging to the solid solution 
from those in L1$_2$ precipitates are the same as the ones developed 
in our study of the precipitation kinetics in the two binary Al-Zr and Al-Sc alloys:
perfect order is assumed for L1$_2$ cluster\footnote{Zr and Sc atoms
belong to a L1$_2$ cluster if all their first nearest neighbors are 
Al atoms and at least one of their six nearest neighbors is a Zr or Sc atom.} 
and a critical size $n_{\mathrm{X}}^*$ is defined so as to differentiate 
unstable clusters from precipitates.

\subsubsection{Precipitation kinetics in the ternary alloy.}
\begin{figure}[!hbp]
	\hfill
	\begin{minipage}{0.3\textwidth}
		\begin{center}
		\begin{tabular}{cc}
		\includegraphics[width=0.5cm]{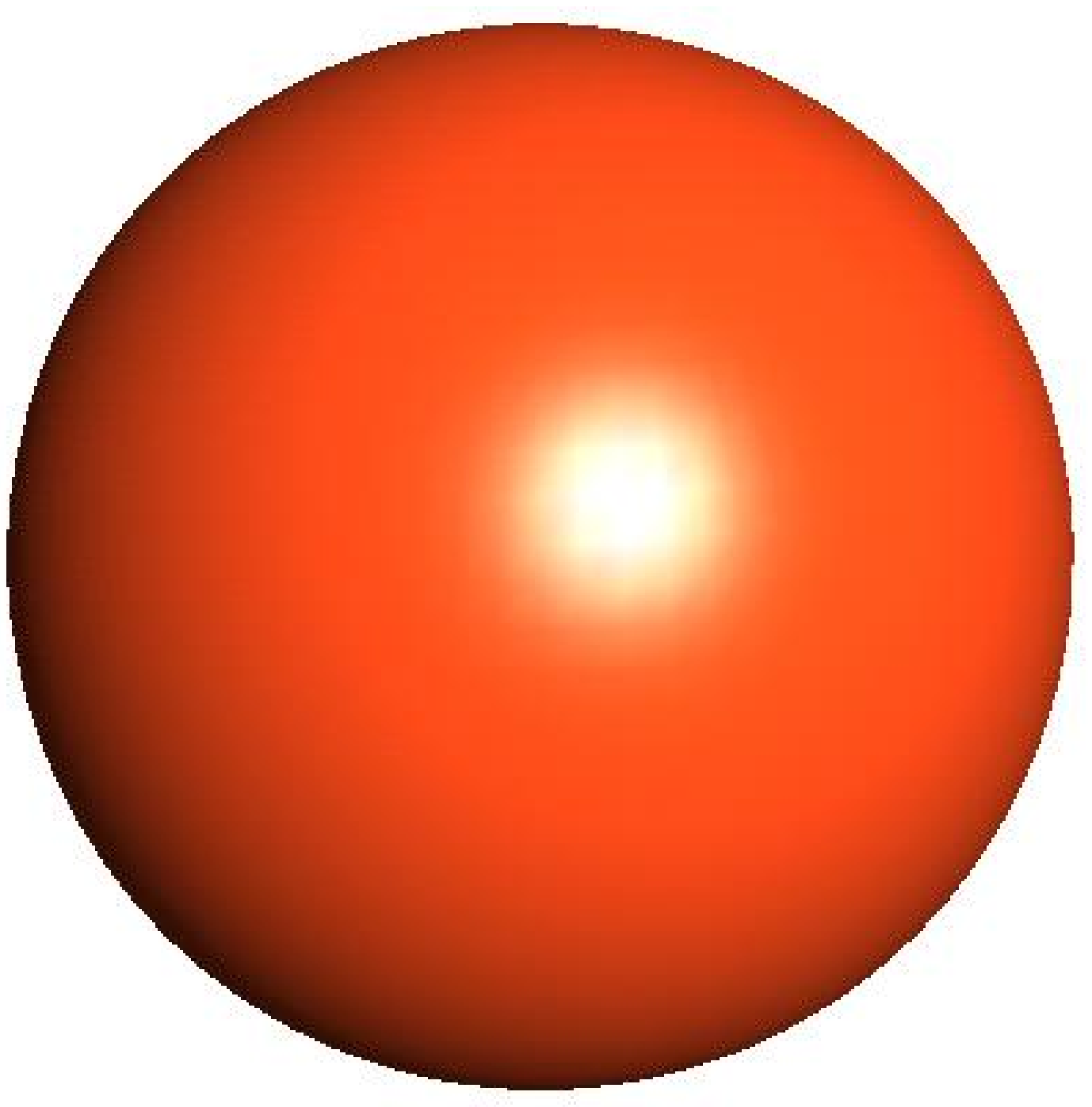} & Sc \\
		\includegraphics[width=0.5cm]{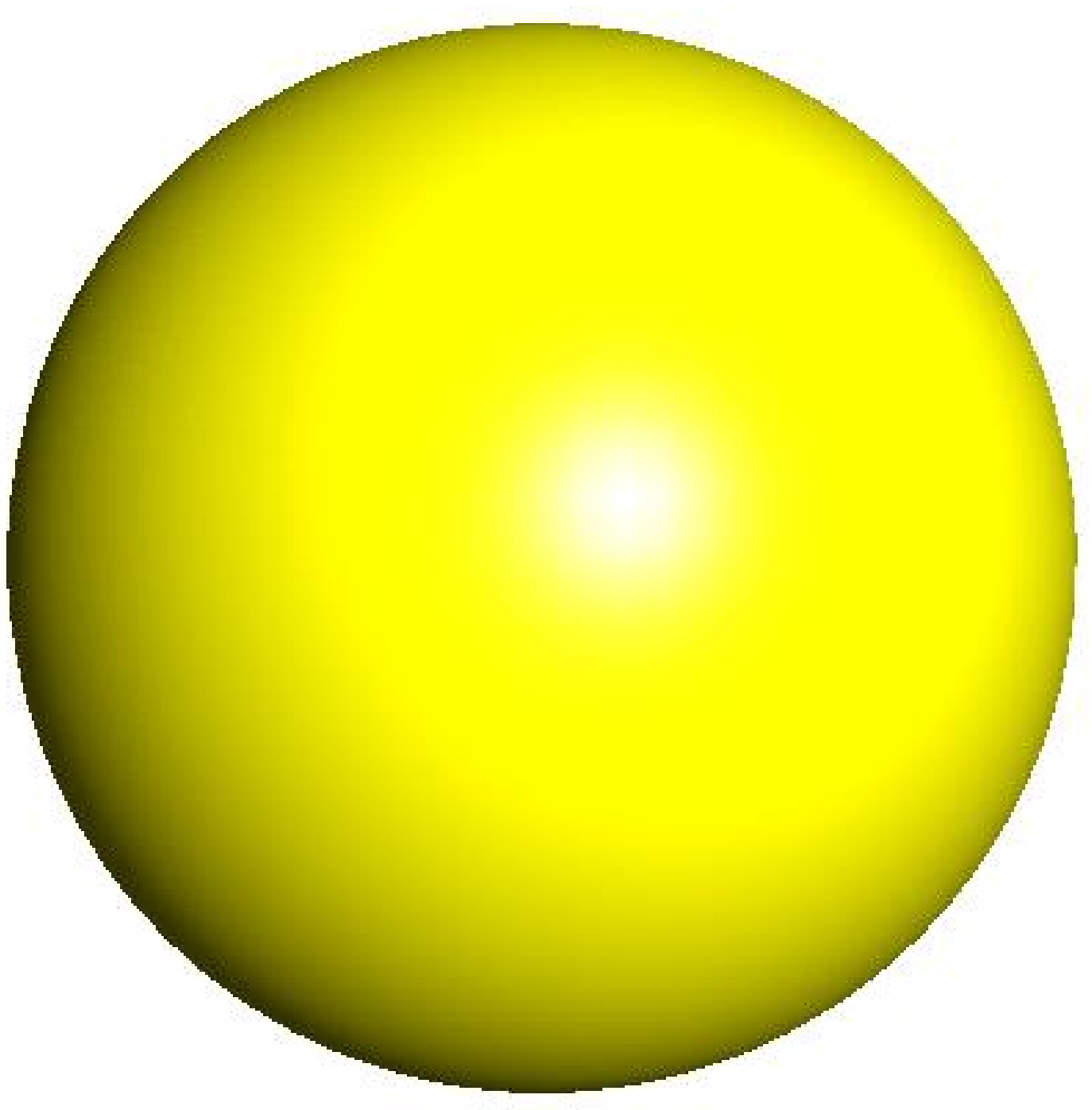} & Zr \\
		\end{tabular}

		\includegraphics[width=0.99\textwidth]{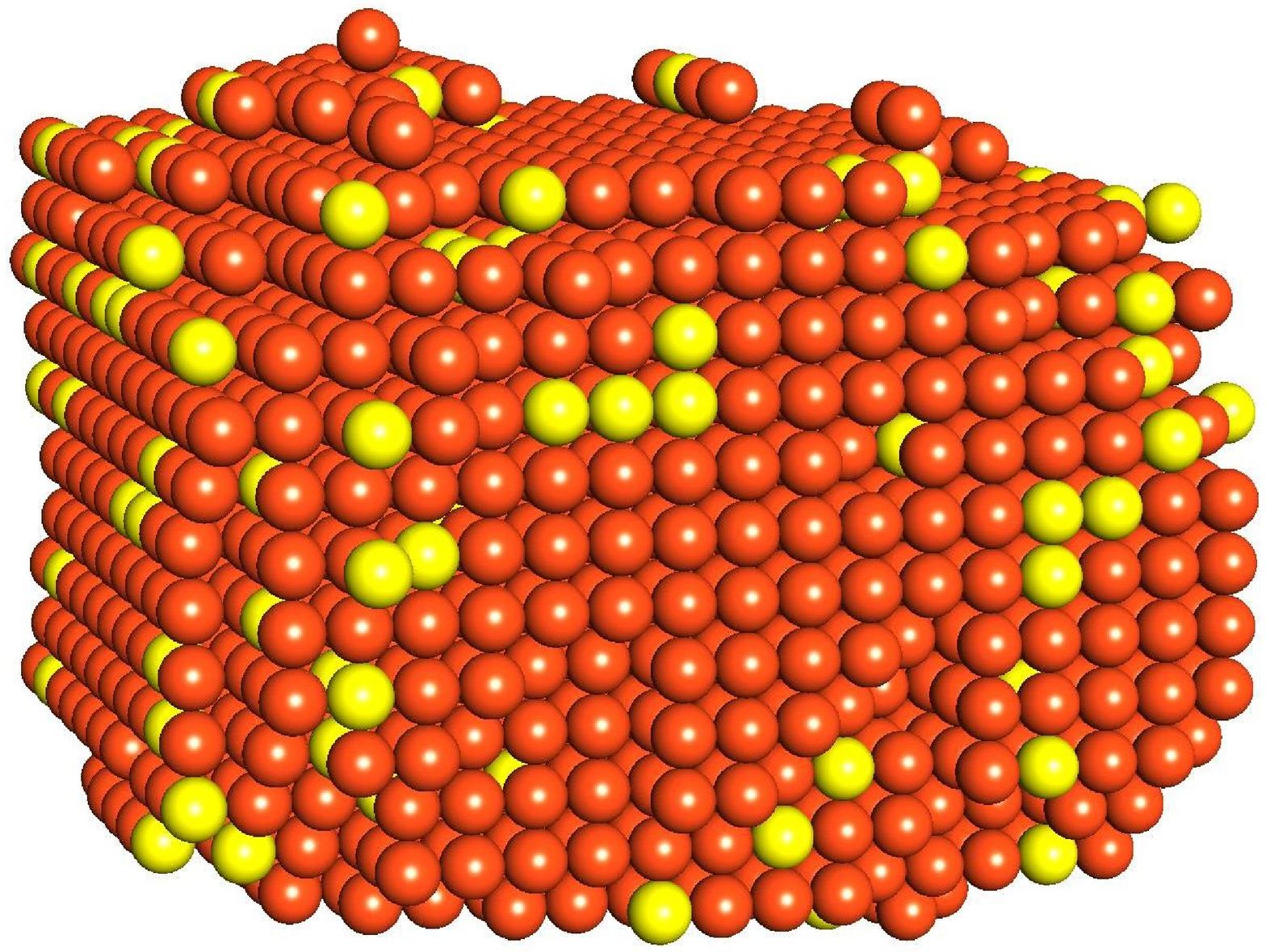}
		\end{center}
	\end{minipage}
	\hfill
	\begin{minipage}{0.5\textwidth}
		\includegraphics[width=0.99\textwidth]{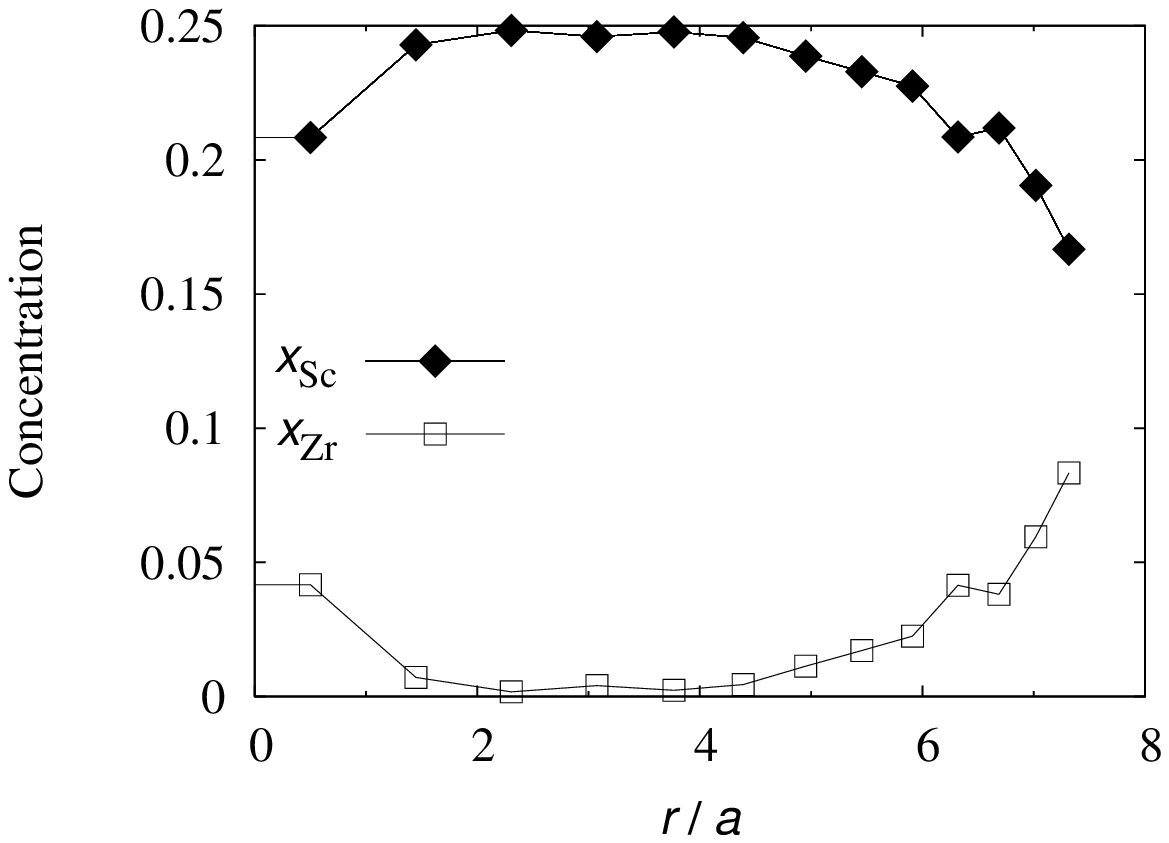}
	\end{minipage}
	\hfill
\caption{Precipitate obtained after the annealing during 1.2~s at 550°C
of an aluminum solid solution containing 0.1~at.\% Zr and 0.5~at.\% Sc
simulated with KMC (for clarity, only Zr and Sc atoms are shown). 
The corresponding radial concentration profile shows the Zr enrichment
of the periphery compared to the core of the precipitate.}
\label{fig:proxi_0.1Zr_0.5Sc_550C}
\end{figure}

Using this atomic diffusion model, we run KMC simulations
of the annealing of supersaturated Al-Zr-Sc solid solutions.
Precipitates appearing in the simulation box are Al$_3$Zr$_x$Sc$_{1-x}$
ternary compounds having the partially ordered L1$_2$ structure 
predicted by thermodynamics.
These precipitates are inhomogeneous (Fig.~\ref{fig:proxi_0.1Zr_0.5Sc_550C}):
their core is richer in Sc than in Zr whereas this is the opposite
for the external shells.
Zr concentration slightly decreases away from the core and the intermediate 
shells have almost the Al$_3$Sc stoichiometry.
As for the external shells, they are strongly enriched in Zr.
This structure predicted by our atomic model agrees with experimental observations
recently made with high resolution electron microscopy \cite{TOL05,RAD05} 
and with three-dimensional atom probe \cite{FOR04,FUL05a}.

\begin{figure}[!hbt]
	\begin{center}
		\includegraphics[width=0.5\textwidth]{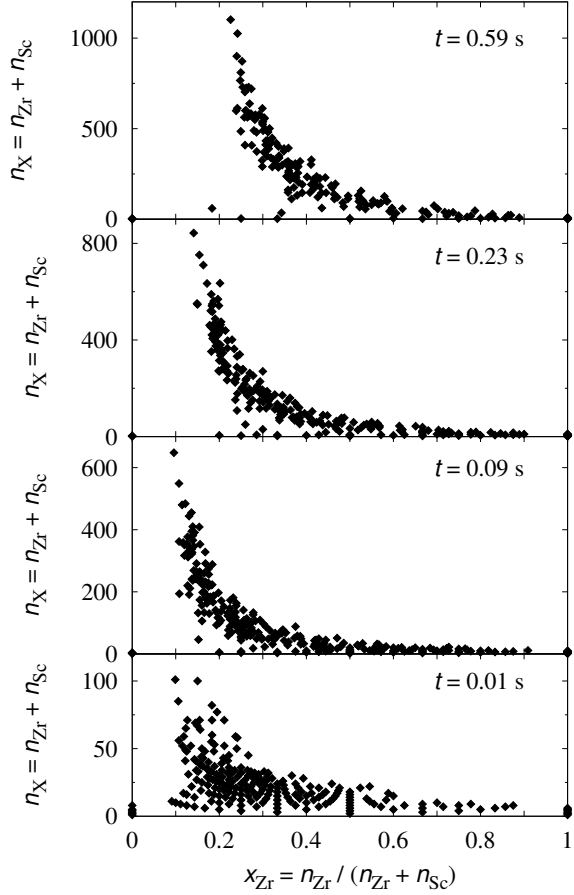}
	\end{center}
	\caption{Projection in the plane $x_{\mathrm{Zr}}$, $n_{\mathrm{X}}$
	of the precipitate distribution in a solid solution of composition 
	$x^0_{\mathrm{Zr}}=0.5$~at.\% and $x^0_{\mathrm{Sc}}=0.5$~at.\%
	annealed at $T=550^{\circ}$C.}
	\label{fig:projection_0.5Zr_0.5Sc_550C}
\end{figure}

Following the composition and the size of all clusters present in the simulation box
(Fig.~\ref{fig:projection_0.5Zr_0.5Sc_550C}), we see that small clusters, which should 
be sub-critical, can have all compositions between the two Al$_3$Zr and Al$_3$Sc
stoichiometric compounds. 
As for over-critical clusters, the smaller ones
are richer in Zr than the larger ones. These small clusters, which have just nucleated,
grow by absorbing Sc atoms, leading to the tail of the distribution 
in the Sc richer part.
As there is no diffusion inside precipitates (because of a high vacancy energy of formation
and of the energetic cost associated with the creation of antisite defects),
Zr and Sc atoms cannot homogenize.
Therefore this distribution is consistent with the previous observation that the 
core of the precipitates can contain Zr whereas the intermediate shells 
contain almost exclusively Sc. 
This shows that Zr plays a role during nucleation but 
that growth is controlled by Sc, 
because this element diffuses much faster than Zr in the matrix.
As precipitation is going on, the solid solution depletes in Sc.
Therefore clusters can only grow by absorbing Zr atoms, leading 
to a move of the whole distribution towards the Zr richer part 
(Fig.~\ref{fig:projection_0.5Zr_0.5Sc_550C}).
This is consistent with the fact that the external shells of the precipitates 
are richer in Zr. 
The precipitate growth by Zr absorption occurring in a second stage 
leads to a slowdown of the precipitation kinetics.

\begin{figure}[!hbt]
	\begin{center}
		\includegraphics[width=0.49\linewidth]{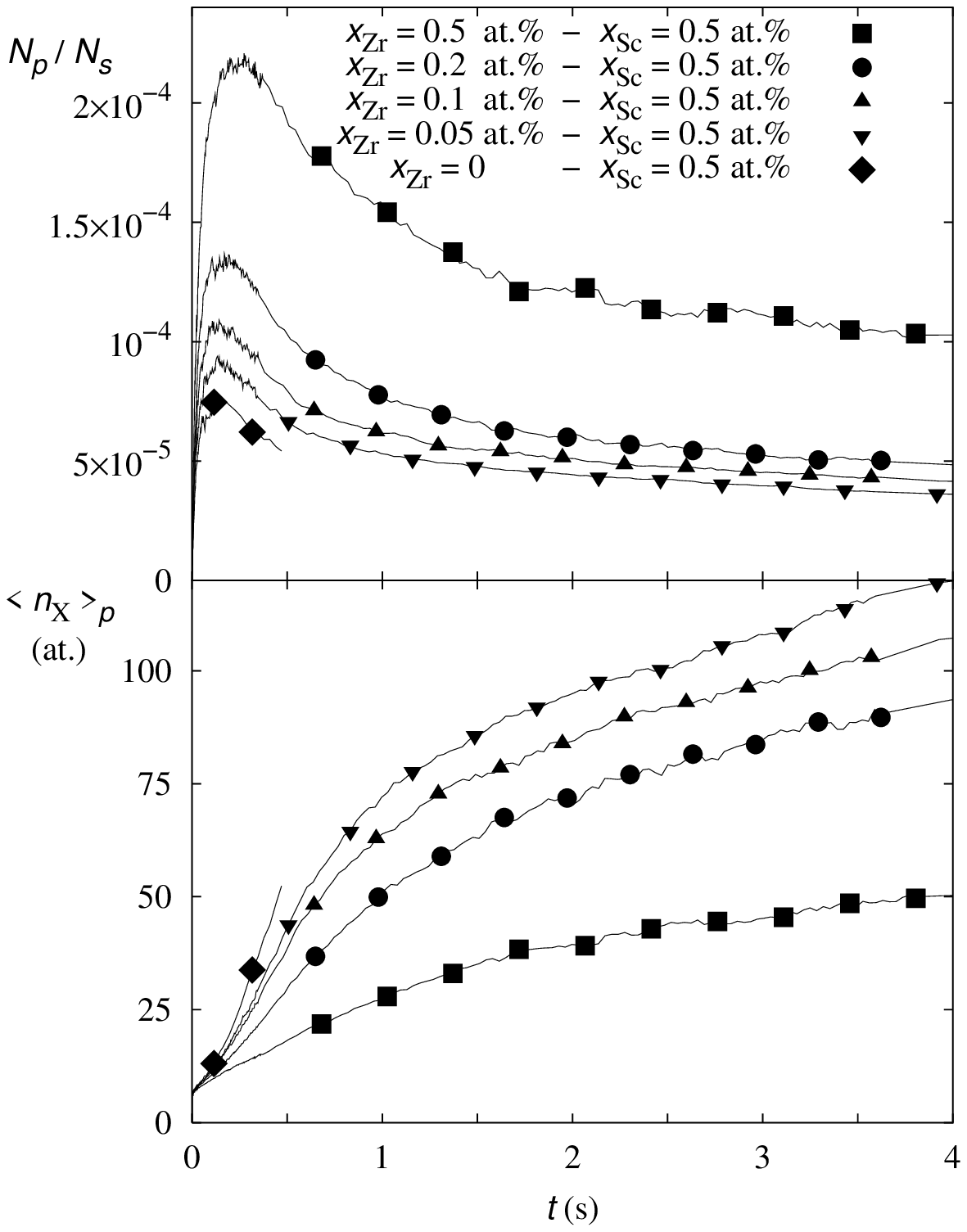}\hfill
		\includegraphics[width=0.49\linewidth]{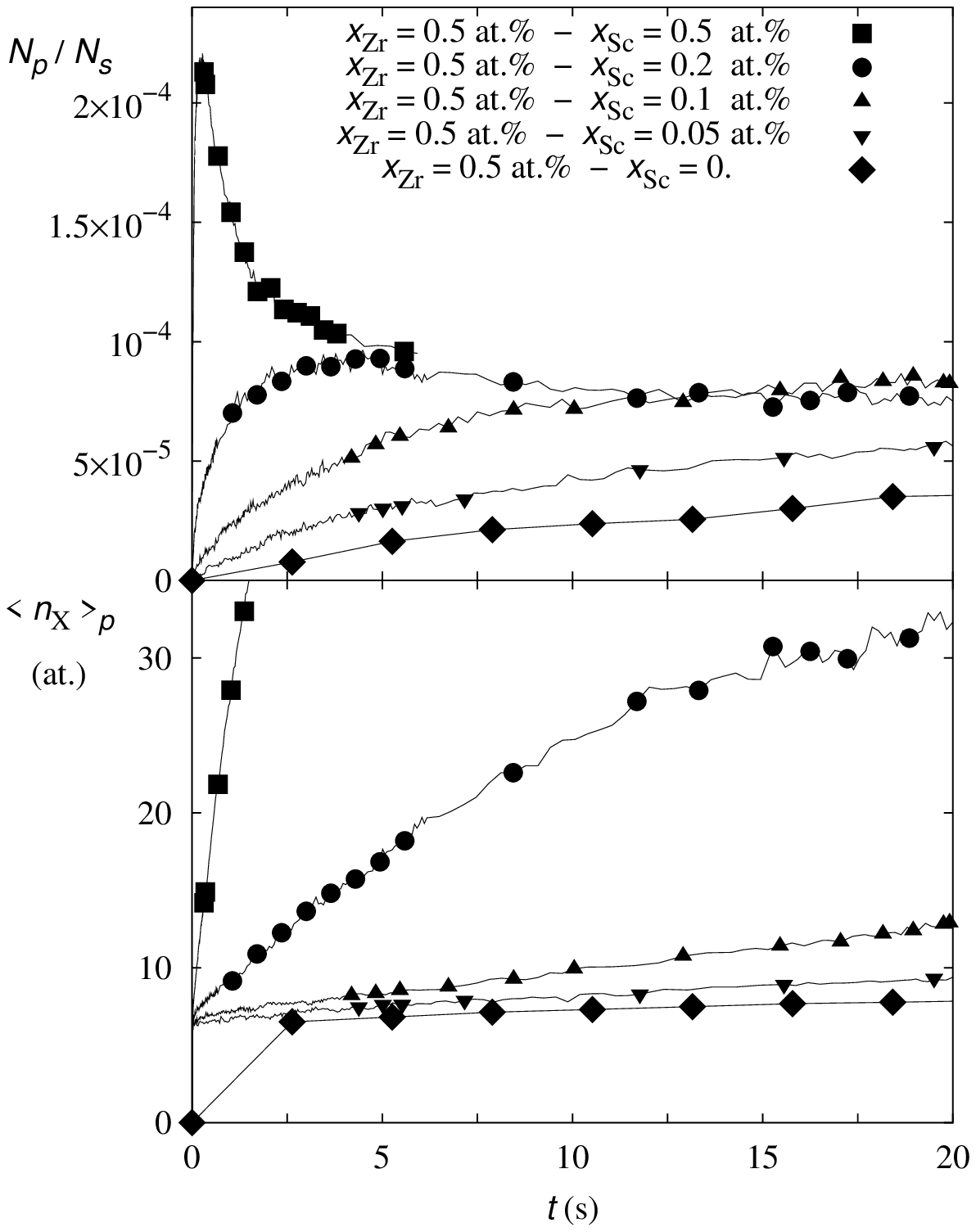}\hfill
	\end{center}
	\caption{Kinetics of precipitation at $T=450^{\circ}$C:
	variation with the Zr and Sc concentration of the number $N_p$ of precipitates 
	and of their mean size $< n_{\mathrm{X}} >$. 
	The critical size to discriminate precipitates from sub-critical clusters is
	$n^*_{\mathrm{X}}=6$.} 
	\label{fig:cinetique_450C1}
\end{figure}

In agreement with TEM observations \cite{YEL85,FUL03}, these atomic simulations
show that a Zr addition to an Al-Sc alloy leads to a higher density of precipitates, 
these precipitates being smaller (Fig.\ref{fig:cinetique_450C1}). 
Our atomic model allows to understand this effect.
A Zr addition increases the nucleation driving force and thus the number of precipitates.
As these ones mainly grow by absorbing Sc, one obtains smaller precipitates 
because the number of growing clusters is higher and there is less Sc available for 
growth at the end of the nucleation stage. 
The precipitate growth by Zr absorption is really slow but plays an important role too 
as it leads to the formation of an external Zr enriched shell which is responsible 
for the good resistance against coarsening.
This kinetic effect due to the addition of an impurity having an attractive interaction
with solute atoms ($\omega^{(2)}_{\mathrm{ZrSc}}<0$) has been already noticed 
by Soisson and Martin \cite{SOI00}.

It is interesting to notice that a Sc addition to an Al-Zr alloy has quite distinct effects
(Fig.~\ref{fig:cinetique_450C1}).
As Sc atoms plays a key role for nucleation and growth, such an addition leads to 
an increase of the precipitate density as well as to an increase of the precipitate size.
Indeed the nucleation stage is shorter and the number of nucleating precipitates 
is higher. 
But this increase of the nuclei density does not consume the whole Sc in the solid solution
and at the end of the nucleation stage, enough Sc remains so as to accelerate
precipitate growth and coarsening, thus leading to bigger precipitates.
In this case, one has to consider not only the thermodynamic contribution of 
the impurity addition, but its kinetic contribution too so as to conclude
on its effects on the precipitation kinetics.

\section{Cluster dynamics (CD)}

KMC simulations allow us to understand precipitation 
kinetics in the ternary Al-Zr-Sc alloy in great details. Nevertheless, these simulations
are restricted to short annealing times and high enough
supersaturations, preventing any comparison with experimental data.
Starting from the same atomic model, we build a mesoscopic modeling \cite{CLO05}
which is based on a set of rate equations describing in the same framework
the three different stages of precipitation, \ie nucleation, growth and coarsening.
It requires only a limited number of parameters, the interface free energy
and the solute diffusion coefficients, which can be quite easily deduced 
from atomic parameters.
The range of supersaturations and annealing times that can be simulated 
are thus extended for both binary Al-Zr and Al-Sc systems
allowing a comparison with experimental data.

\subsection{Mesoscopic model}

In its strict sense, CD rests on the description of the alloy
undergoing phase separation as a gas of solute clusters which exchange
solute atoms by single atom diffusion\cite{GOL95,MAT97,BAR04}. 
Clusters are assumed to be spherical and 
are described by a single parameter, their size or the number 
$\nX$ of solute atoms $\mathrm{X}$ they contain.
In such a description, there is no precisely defined distinction
between the solid solution on the one hand and the precipitates on the other hand,
at variance with the CNT: the distribution of cluster sizes
is the only quantity of interest (for a detailed discussion see Ref. \citen{MAR05}).

\subsubsection{Master equation.}
The CD technique describes the precipitation kinetics
thanks to a master equation giving the time evolution 
of the cluster size distribution\cite{GOL95,MAT97,BAR04}. 
When only monomers can migrate, which is the case for Al-Zr
as well as Al-Sc systems \cite{CLO04}, the probability $C_{\nX}$ 
to observe a cluster containing $\nX$ solute atoms obeys
the differential equations
\begin{subequations}
\begin{align}
\frac{\ud C_{\nX}}{\ud t} &= J_{\nX-1\to\nX} - J_{\nX\to\nX+1} ,\quad \forall\ \nX\geq2 \\
\frac{\ud C_{1}}{\ud t} &= - 2 J_{1\to2} - \sum_{\nX\geq2}{J_{\nX\to\nX+1}},
\end{align}
\label{eq:DA}
\end{subequations}
where the flux $J_{{\nX} \to {\nX}+1}$ from the class of size $\nX$ to the class $\nX+1$
is written
\begin{equation}
J_{{\nX} \to {\nX}+1} = \beta_{\nX} C_{\nX} - \alpha_{{\nX}+1} C_{{\nX}+1}
\label{eq:flux_DA} ,
\end{equation}
$\beta_{\nX}$ being the probability per unit time for one solute atom 
to impinge on a cluster of size $\nX$
and $\alpha_{\nX}$ for one atom to leave a cluster of size $\nX$.

\subsubsection{Condensation rate.}
When the solute long-range diffusion controls the precipitation kinetics,
the condensation rate is obtained by solving the diffusion problem in the solid 
solution around a spherical precipitate.
For a cluster of radius $r_{\nX}$, this condensation rate takes the form \cite{WAI58,MAR78}
\begin{equation}
\beta_{\nX} = 4 \pi r_{\nX} \frac{D_{\mathrm{X}}}{\Omega} C_1,
\label{eq:beta_n}
\end{equation}
where $D_{\mathrm{X}}$ is the diffusion coefficient of the solute at infinite 
dilution of the solid solution 
and $\Omega$ is the mean atomic volume corresponding to one lattice site.

\subsubsection{Evaporation rate.}
The evaporation rate is obtained assuming that  
it is an intrinsic property of the cluster and
therefore does not depend on the solid solution 
surrounding the cluster.
This means that the cluster has enough time to explore all its configurations
between the arrival and the departure of a solute atom.
Thus $\alpha_{\nX}$ should not depend on the nominal concentration 
of the solid solution and can be obtained by considering 
any undersaturated solid solution of nominal concentration $x^0_{\mathrm{X}}$.
Such a solid solution is stable. Then there should be no energy dissipation.
This involves that all fluxes $J_{\nX\to\nX+1}$ equal zero.
Using equation \ref{eq:flux_DA}, one obtains
\begin{equation}
\alpha_{{\nX}+1} = \bar{\alpha}_{{\nX}+1}(x^{0}_{\mathrm{X}}) 
= \bar{\beta}_{\nX}(x^{0}_{\mathrm{X}}) 
\frac{\bar{C}_{\nX}(x^{0}_{\mathrm{X}})}{\bar{C}_{{\nX}+1}(x^{0}_{\mathrm{X}})}, 
\label{eq:an_equilibre}
\end{equation}
where overlined quantities are evaluated in the solid solution at equilibrium.
This finally leads to
\begin{equation}
\alpha_{{\nX}+1} =  4 \pi r_{\nX} \frac{D_{\mathrm{X}}}{\Omega} 
\exp{\left[ (G_{\nX+1} - G_{\nX} - G_{1} ) / kT\right]},
\label{eq:an}
\end{equation}
where $G_{\nX}$ is the free energy of a cluster containing $\nX$ atoms.
It can be divided into a volume and an interface contributions, 
which involves the following expression of the evaporation rate:
\begin{equation}
\alpha_{{\nX}+1} =  4 \pi r_{\nX} \frac{D_{\mathrm{X}}}{\Omega} 
\exp{\left[ (36\pi)^{1/3} a^2 \left(
(\nX+1)^{2/3} \sigma_{\nX+1} - {\nX}^{2/3} \sigma_{\nX} - \sigma_{1} \right) / kT\right]},
\label{eq:an_sigma}
\end{equation}
where $\sigma_{n_{\mathrm{X}}}$ is the interface free energy of a cluster
containing $\nX$ solute atoms and $a$ the lattice parameter.\footnote{A more 
detailed description of the derivation of Eq. \ref{eq:an_sigma} is given
in Ref. \citen{CLO05}.}

Looking at the expression \ref{eq:beta_n} of the condensation rate 
and the expression \ref{eq:an_sigma} of the evaporation rate, 
one sees that the only parameters needed by CD are the diffusion
coefficient and the interface free energy.
There is no need to know the nucleation free energy in opposition to 
other mesoscopic models based on CNT.
In CD, thermodynamics of the solid solution is described thanks to a lattice 
gas model and therefore the nucleation free energy results from this description 
and is not an input of the modeling.

\subsubsection{Interface free energy.}
The interface free energy can be deduced from the atomic model \cite{CLO04}
by computing within the Bragg-Williams approximation
the free energies corresponding to planar interfaces
between the aluminum solid solution and the L1$_2$ precipitates
for the three most dense packing orientations \{111\}, \{110\} and \{100\}.
We obtain $\sigma_{100} < \sigma_{110} < \sigma_{111}$,
indicating that precipitates mainly
show facets in the \{100\} direction and that facets in the 
\{110\} and \{111\} directions are small.
As the difference between $\sigma_{100}$, $\sigma_{110}$ and
$\sigma_{111}$ is decreasing with temperature, precipitates are
becoming more isotropic at higher temperatures.
This is in agreement with the shapes of the precipitates observed
during the atomic simulations, as well as with experimental 
observations done by Marquis and Seidman \cite{MAR01}.

So as to get the input parameters needed by CD,
one has to deduce an isotropic average interface free energy $\bar{\sigma}$ 
from these planar interface free energies.
This can be done using the Wulff construction \cite{PORTER,CHRISTIAN}:
$\bar{\sigma}$ is defined so as to give the same interface free energy 
for a spherical precipitate having the same volume as the real faceted
one.\footnote{Details of the calculation can be found in the appendix
A of Ref.\citen{CLO04}.}
For temperatures ranging between 0~K and the aluminum melting temperature
(933.5~K), $\bar{\sigma}$ is varying between 124 and 92~mJ.m${^{-2}}$ for Al$_3$Zr precipitates
and between 138 and 95~mJ.m${^{-2}}$ for Al$_3$Sc precipitates.

\begin{figure}[!hbt]
	\begin{center}
		\includegraphics[width=0.6\linewidth]{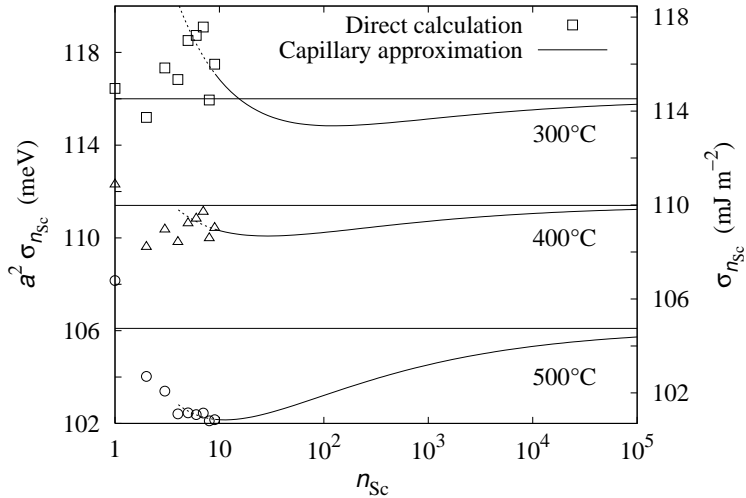}
	\end{center}
	\caption{Variation with the cluster size $\nSc$ of the interface
	free energy between the solid solution and Al$_3$Sc at three different
	temperatures (300, 400, and 500$^{\circ}$C). 
	Symbols correspond to $\sigma_{\nSc}$ as given by the direct calculation
	of the cluster formation free energy (Eq. 20 of Ref. \citen{CLO04})
	and lines to the capillary approximation (Eq.~\ref{eq:capillary_sigma})
	as well as to its asymptotic limit.
	Dotted lines indicate the range of sizes where the coefficients
	$c$ and $d$ have been fitted.}
	\label{fig:sigma}
\end{figure}

The interface free energy $\bar{\sigma}$ does not depend on the size of the cluster
and corresponds to the asymptotic limit of the quantity $\sigma_{\nX}$ 
used to define the evaporation rate (Eq.~\ref{eq:an_sigma}). 
A direct calculation considering the partition functions of small clusters \cite{CLO04} shows that 
$\sigma_{\nX}$ slightly deviates from this asymptotic value at small sizes (Fig.~\ref{fig:sigma}).
Nevertheless, CD is very sensitive to this input parameter because
its thermodynamic description only relies on it.
Therefore, the size dependence of $\sigma_{\nX}$ has to be considered.
In this purpose, for clusters containing no more than 9 solute atoms, 
the interface free energy is computed precisely using
Eq. 20 of Ref. \citen{CLO04} 
and for clusters of size $\nX\geq10$, the interface free energy is obtained using
an extension of the capillary approximation,
\begin{equation}
\sigma_{\nX} = \bar{\sigma}\left( 1 + c\ \nX^{-1/3} + d\ \nX^{-2/3} \right),
\label{eq:capillary_sigma}
\end{equation}
where $c$ and $d$ respectively correspond to the \emph{line} and \emph{point} contributions 
\cite{PER84}.
The asymptotic value $\bar{\sigma}$ used is the one previously calculated 
whereas coefficients $c$ and $d$ are obtained by a least square fit of the 
exact expression for sizes $5\geq\nX\geq9$.

\subsection{Precipitation kinetics}

\begin{figure}[!hbt]
	\begin{center}
		\includegraphics[width=0.47\linewidth]{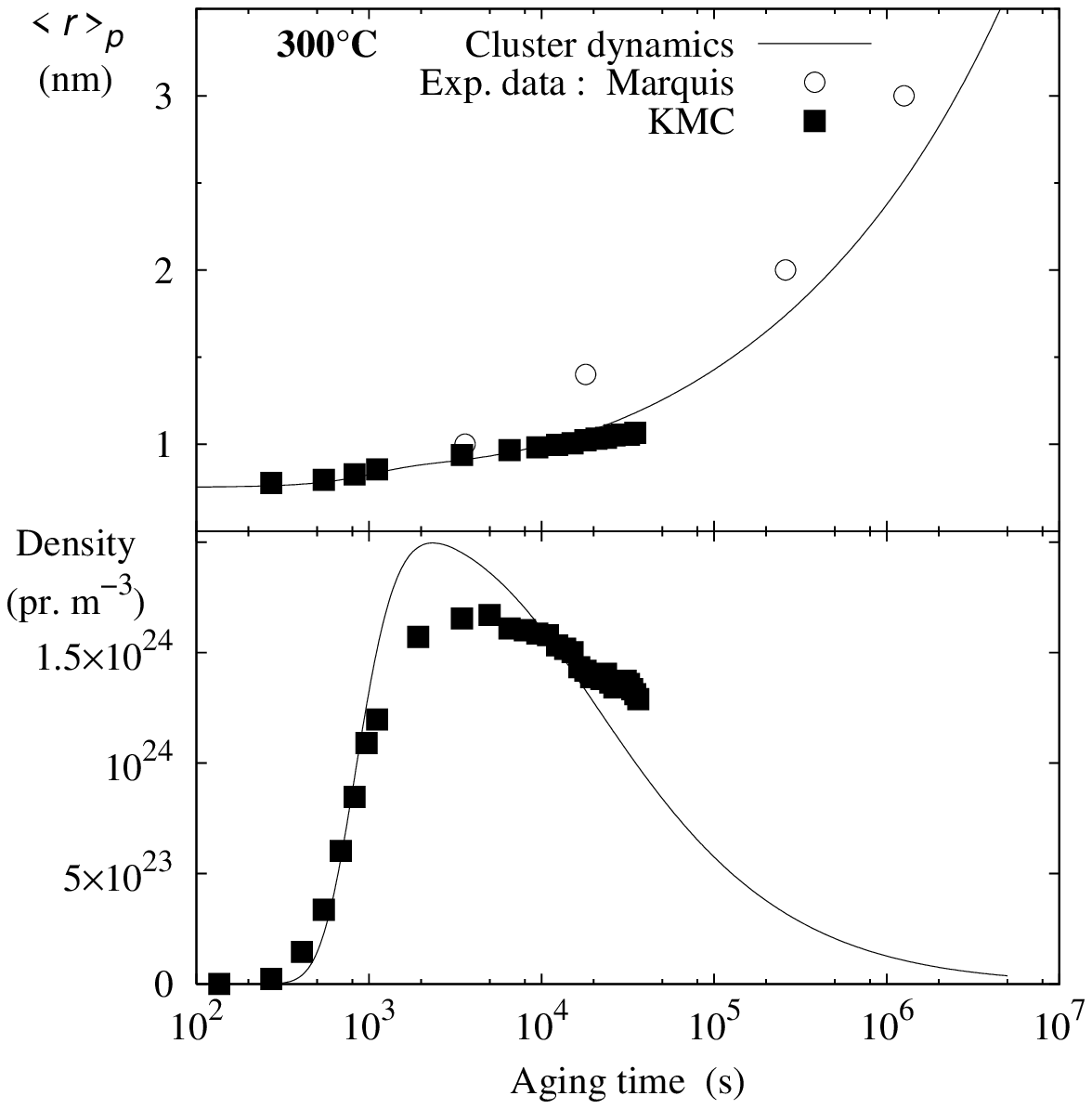}
		\hfill
		\includegraphics[width=0.47\linewidth]{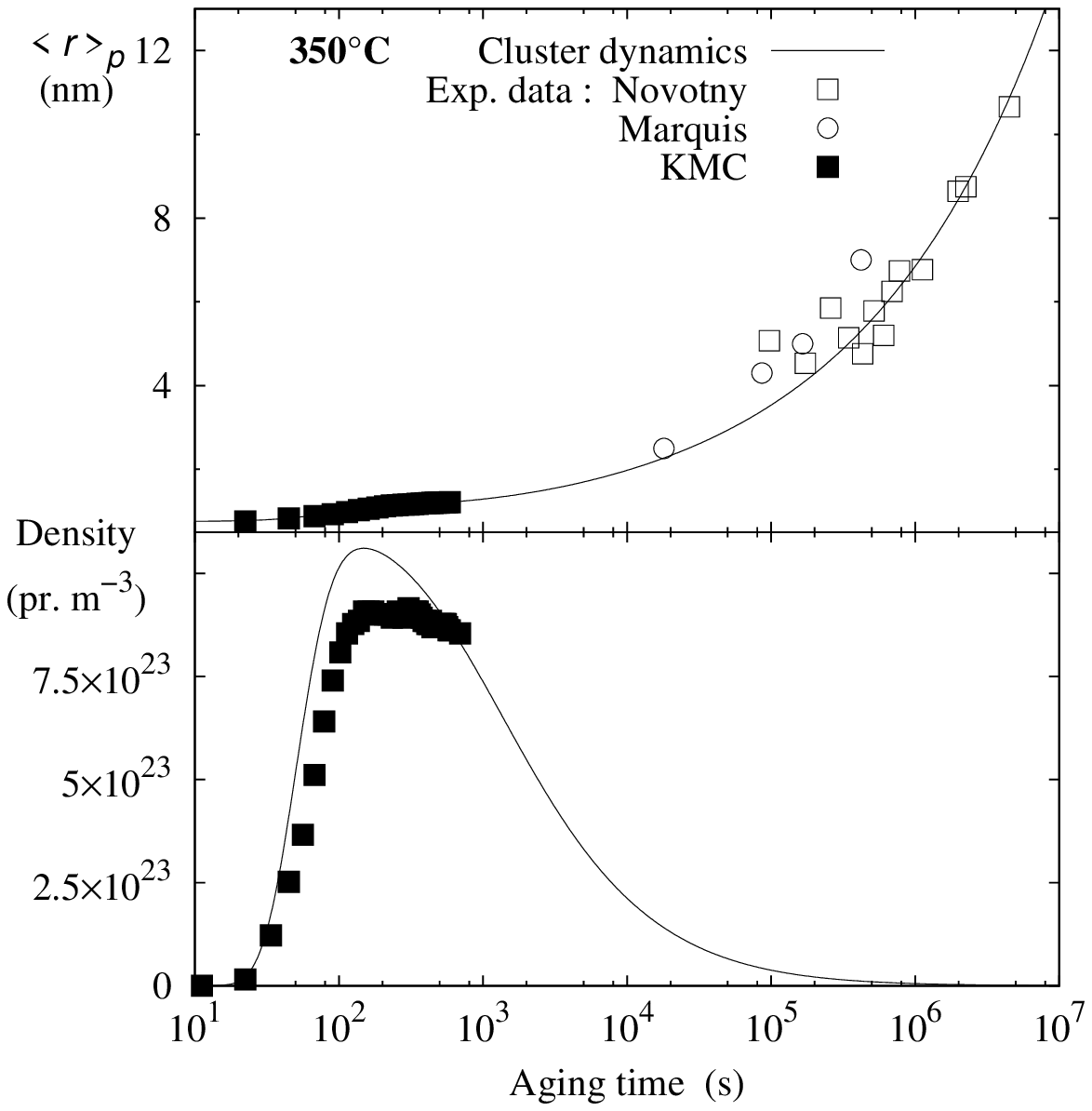}
		\hfill
		\includegraphics[width=0.47\linewidth]{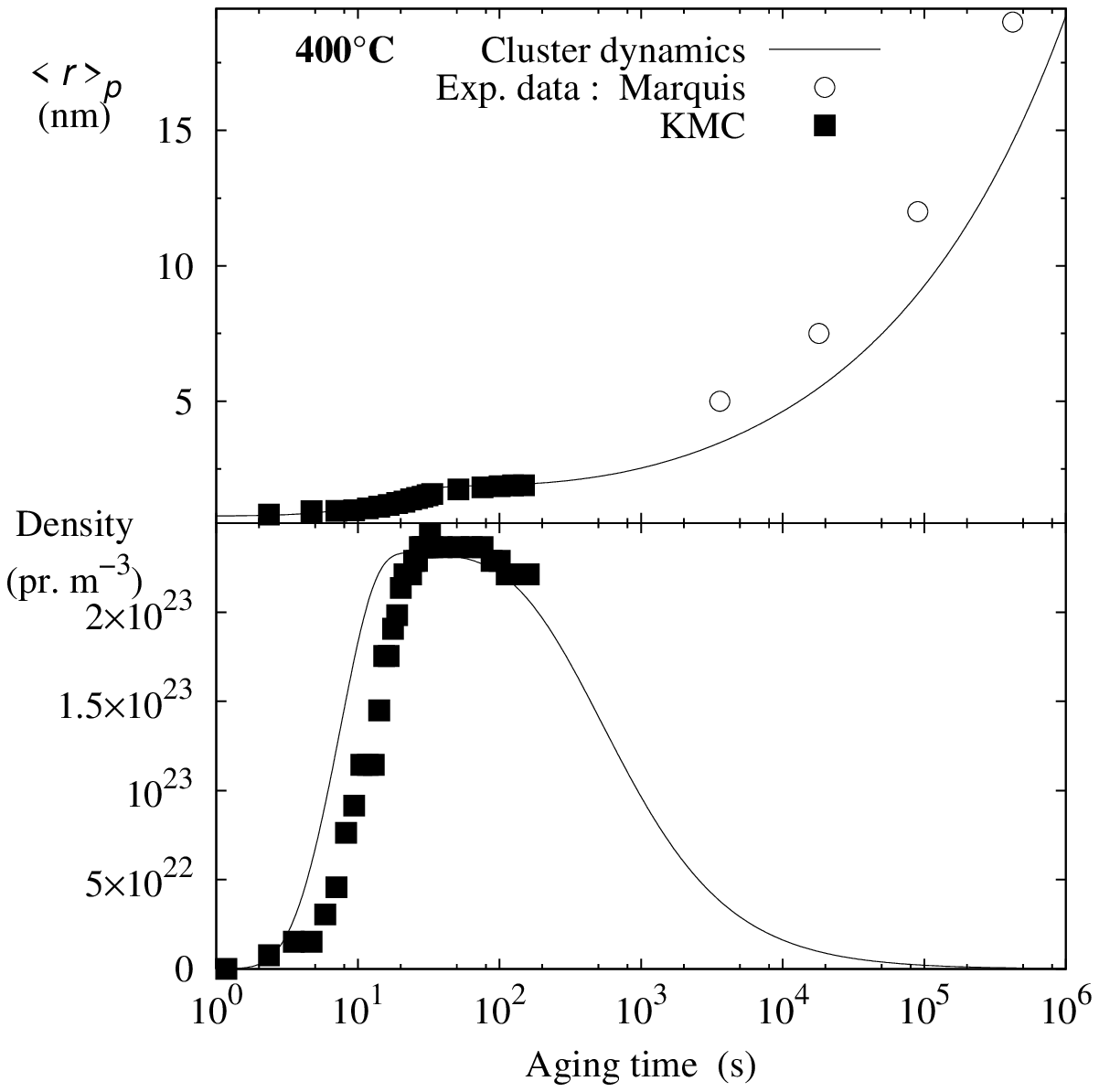}
		\hfill
	\end{center}
	\caption{Mean precipitate radius and precipitate density
	as a function of the aging time for a solid solution 
	of composition $x^0_{\mathrm{Sc}}=0.18$~at.\% 
	at different temperatures ($T=300$, 350, and 400°C) 
	as given by CD and compared to experimental
	data\cite{NOV01,MAR01,MAR02,MAR02T} 
	and to KMC results\cite{CLO04}.
	The cutoff radius used for CD and KMC is 
	$r^*\sim0.75$~nm ($\nX^*=27$).}
	\label{fig:cinetique_da1}
\end{figure}

In order to compare precipitation kinetics obtained from KMC
and from CD simulations,
an arbitrary threshold size $\nX^{*}$ is used to define precipitates
so as to be able to calculate mean quantities like the precipitate size 
and density.
Below this threshold size, clusters are assumed to be invisible. 
When comparing with experimental data, we choose $\nX^{*}=27$ corresponding
to a precipitate diameter equal to 1.5~nm which is supposed to be the smallest
size which can be discriminated by TEM.
This threshold size can thus be different from the critical size
of CNT: the only important thing is 
to use the same criterion to define precipitates when comparing 
atomic and mesoscopic simulations.

Comparing the precipitation kinetics obtained with KMC
and CD simulations, one sees that CD
manages to reproduce the variations of the precipitate density 
and of their mean size (Fig.~\ref{fig:cinetique_da1}). 
A more thorough comparison \cite{CLO05} shows that CD
catches variations with the supersaturations and the annealing
temperatures of the precipitation kinetics. 
For low supersaturations, the agreement between both modeling 
is really good, whereas for high supersaturations, 
kinetics modeled by CD appear to be slightly too slow 
compared to the ones simulated with KMC.
This delay only corresponds to a constant factor on the time scale 
and in all cases the prediction of the precipitate maximal density 
at the transition between the growth and the coarsening stages is correct.

Precipitation kinetics simulated with CD can be compared too
with the experimental data of Novotny and Ardell \cite{NOV01}
and those of Marquis \etal \cite{MAR01,MAR02,MAR02T} who studied
an aluminum alloy having the composition $x^0_{\mathrm{Sc}}=0.18$~at.\%.
For the three different temperatures $T=300$, $350$ and $400^{\circ}$C,
it appears that the CD equations manage to reproduce
the variation with time of the mean precipitate radius (Fig.~\ref{fig:cinetique_da1}).
Despite the fact that the time scales in real experiments are several order of magnitude
larger than those in KMC, CD, with a single set of parameters,
reproduces atomic simulations at short times 
and gives a safe extrapolation thereof to the range of annealing times
that can be compared with experimental data.
In Ref.~\citen{CLO05}, it was shown too that the precipitate size distributions
simulated with CD agrees really well with the experimentally 
ones measured by Novotny and Ardell \cite{NOV01}.

\begin{figure}[!hbt]
	\begin{center}
		\includegraphics[width=0.47\linewidth]{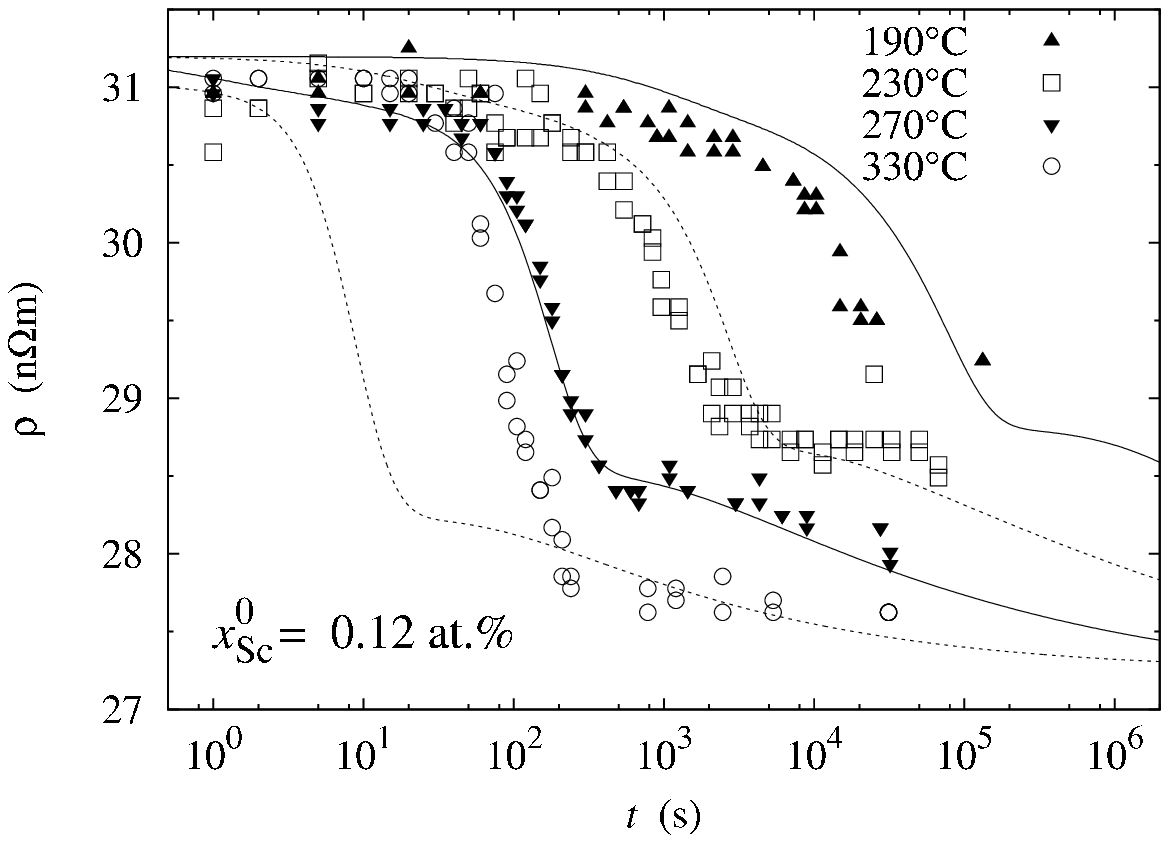}
		\includegraphics[width=0.47\linewidth]{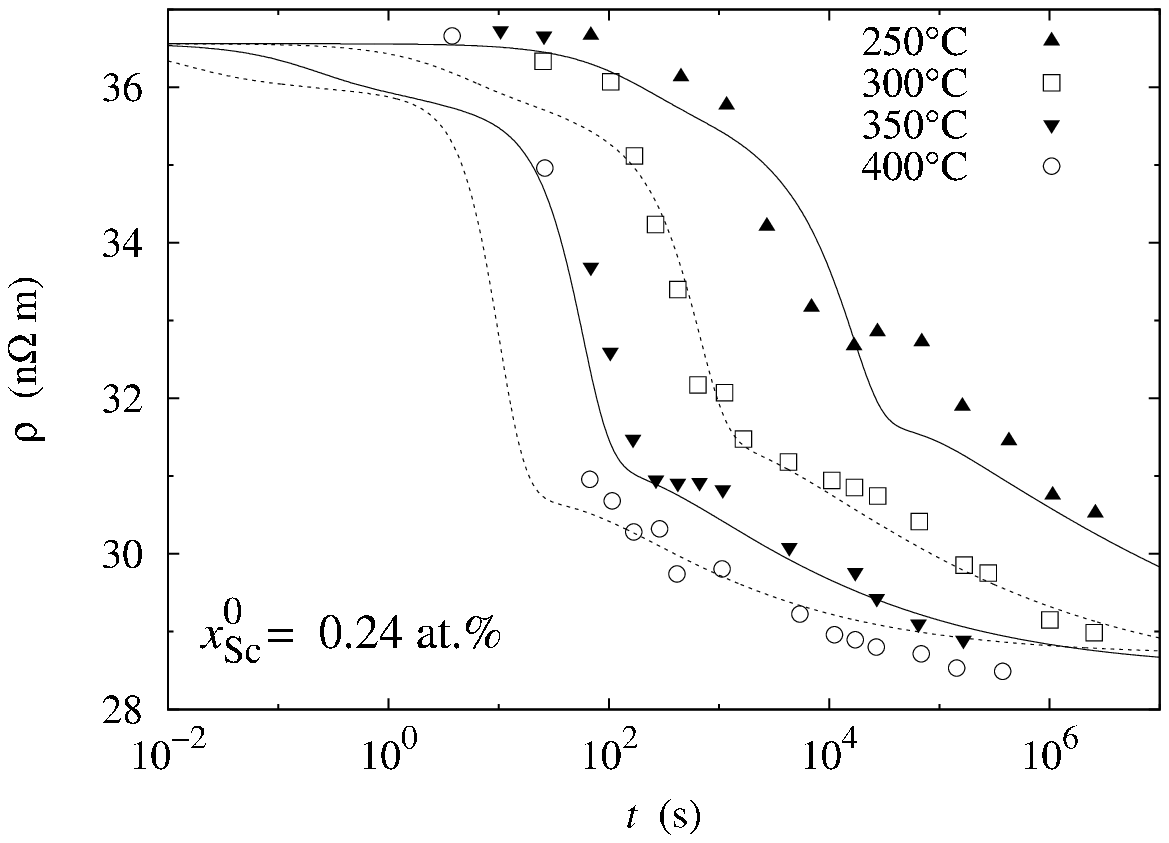}
	\end{center}
	\caption{Time evolution of the resistivity $\rho$ experimentally observed \cite{ZAK97,ROY05c}
	and deduced from CD simulations for two solid solution of composition
	$x^0_{\mathrm{Sc}}=0.12$ and 0.24~at.\%.}
	\label{fig:resistivite}
\end{figure}

CD can be used so as to simulate the evolution of the electric resistivity 
during annealing kinetics. 
Some calculations of the electrical resistivity associated with a cluster distribution
exist in the literature \cite{LUI80,ROS87}, but as a first approximation, one can consider
that the resistivity contribution $\rho_n$ of a cluster of size $n$ is proportional
to its section, $\rho_n=\rho_1n^{2/3}$. 
If one assumes that all clusters are contributing to the resistivity, the resistivity
measured at time $t$ in the phase separating system should be
\begin{equation}
	\rho(t) = \rho^0_{\mathrm{Al}} + \delta\rho_{\mathrm{Sc}}
	\sum_{n=1}^{\infty}{C_n(t) n^{2/3}},
	\label{eq:resistivity_cd}
\end{equation}
where the size distribution $C_n(t)$ is given by CD (Eq.~\ref{eq:DA}).
$\rho^0_{\mathrm{Al}}$ is the resistivity of pure aluminum 
and is temperature dependent.\footnote{Measurements corresponding to 
experimental studies \cite{ZAK97,ROY05c} have been performed at different temperatures.
When comparing with R{\o}yset and Ryum experimental data\cite{ROY05c}, we use 
$\rho^0_{\mathrm{Al}}=27$~n$\Omega$m,
whereas we use $\rho^0_{\mathrm{Al}}=28.4$~n$\Omega$m for the comparison
with Zakharov data \cite{ZAK97}.}
The increase of resistivity with Sc content has been measured at 77~K
by Fujikawa \etal \cite{FUJ79,JO93} who give 
$\delta\rho_{\mathrm{Sc}}=3400$~n$\Omega$m. 
Assuming that Matthienssen rule\cite{FLYNN} is obeyed, this quantity 
does not depend on the temperature.
When comparing resistivity predicted by CD
with the experimental ones measured during precipitation kinetics \cite{ZAK97,ROY05c} 
we obtain a good agreement (Fig.~\ref{fig:resistivite}),
especially for the lowest temperatures: CD manages to reproduce 
the fast decrease of resistivity during the nucleation and growth stages 
as well as the slower variation which follows during the coarsening
stage.
For the highest temperatures (330$^{\circ}$C for the solid solution of 
concentration $x^0_{\mathrm{Sc}}=0.12$~at.\% and 400°C for $x^0_{\mathrm{Sc}}=0.24$~at.\%),
CD appears to be too fast compared to experimental data. 
This may arise from the approximation we are using for the contributions to resistivity
of the various clusters as this size dependence may be too crude for small
clusters. One should notice too that the evolution predicted by CD
at these temperatures is really fast and that the resistivity drop appears at times
($t\sim10$~s) which are too small to be precisely observed experimentally.

CD is thus a powerful modeling technique to study precipitation in a binary alloy, 
allowing to link different scales. 
Starting from an atomic diffusion model, it gives quantitative 
predictions of precipitation kinetics in agreement with experimental data.
It is possible to generalize this mesoscopic technique so as to study precipitation
in a ternary alloy\cite{LAE05} but some work still needs to be done.
In the case of the Al-Zr-Sc alloy, one of the main difficulties is
to find a way for CD to describe the precipitate inhomogeneities. 
Therefore, in the following, we will not use this technique to model 
the whole kinetics of precipitation in the ternary alloy, but we will focus on the nucleation 
stage and see how CNT can be extended so as to predict the nucleation 
rate as well as the compositions $x$ of the Al$_3$Zr$_x$Sc$_{1-x}$ nuclei in this ternary alloy. 

\section{Classical nucleation theory (CNT)}

CNT\footnote{Detailed descriptions
of the theory can be found in Ref.~\citen{MAR78,WU97,WAG91}.}
allows to predict the nucleation rate in a supersaturated 
binary solid solution.
Comparisons with KMC simulations 
\cite{RAM99,SHN99,NOV00,SOI00,BER04,BER04b,CLO04}
show that its predictions are correct as long as 
its input parameters are carefully evaluated.
In particular, for the  Al-Zr and Al-Sc alloys,
it has been shown that one has to take into account
the ordering tendency of the binary alloy when calculating the 
nucleation driving force\cite{CLO04}.
We will see how this can be done in a convenient way as well as
how CNT applies to the two binary
Al-Zr and Al-Sc alloys, before extending the theory
to the ternary Al-Zr-Sc alloy.

\subsection{Binary Al-Zr and Al-Sc alloys}

Whereas CD describes the time evolution of the 
whole cluster size distribution (Eq.~\ref{eq:DA}), 
CNT assumes that the distribution is stationary
in the solid solution, \ie below a critical size $\nX^*$, 
and does not provide any information on the distribution of larger clusters.
The probability to observe a cluster containing $\nX$ solute atoms
for $\nX\leq\nX^*$ is thus given by
\begin{equation}
	\label{eq:distri_cnt}
	C_{\nX} = \exp{\left( - \Delta G_{\nX} / kT \right)},
\end{equation}
where $\Delta G_{\nX}$ is the formation free energy of the 
cluster.\footnote{The formation free energy can be related 
to the free energy $G_{\nX}$ of the cluster by
$\Delta G_{\nX}(\xO)=G_{\nX}-2\nX\mu(\xO)$,
where $\mu(\xO) = \left( \mu_{\mathrm{X}}(\xO) -
\mu_{\mathrm{Al}}(\xO) \right)/2$ is the effective
potential, \ie a Lagrange multiplier imposing that the nominal
concentration of the solid solution is $\xO$.}
It can be evaluated thanks to the capillary approximation, 
dividing this formation free energy into a volume and a surface
contribution,
\begin{equation}
	\label{eq:Gn_cnt}
	\Delta G_{\nX}(\xO) = 4\nX \Delta G^{nuc}(\xO)
	+ \left(36\pi\right)^{1/3} {\nX}^{2/3} a^2 \sigma_{\nX}.
\end{equation}
One thus sees that CNT needs one more input
parameter than CD, the nucleation free energy $\Delta G^{nuc}(\xO)$.

With this additional parameter and the assumption on the cluster
size distribution in the metastable solid solution, CNT
leads to the following prediction of the steady-state 
nucleation rate,
\begin{equation}
	\label{eq:Jst_cnt}
	J^{st} = 4 N_s \beta_{\nX^*}
	\sqrt{ -\frac{1}{2\pi kT} \left.\frac{\partial^2{\Delta G_{\nX}}}{\partial{\nX}^2} \right|_{\nX=\nX^*}}
	\exp{ \left(- \frac{\Delta G_{\nX*}}{kT} \right)},
\end{equation}
where $N_s$ is the number of lattice site. 
The condensation rate $\beta_{\nX^*}$ takes the same expression\footnote{The monomer
concentration $C_1$ appearing in Eq. \ref{eq:beta_n} has to be replaced 
by the nominal concentration $x^0_{\mathrm{X}}$ in CNT.} 
as in CD (Eq.~\ref{eq:beta_n}) and,
like the formation free energy and its second derivative,
it is evaluated for the critical size $\nX^*$
corresponding to the maximum of the formation free energy.

\begin{figure}[btp]
	\begin{center}
		\includegraphics[width=0.49\linewidth]{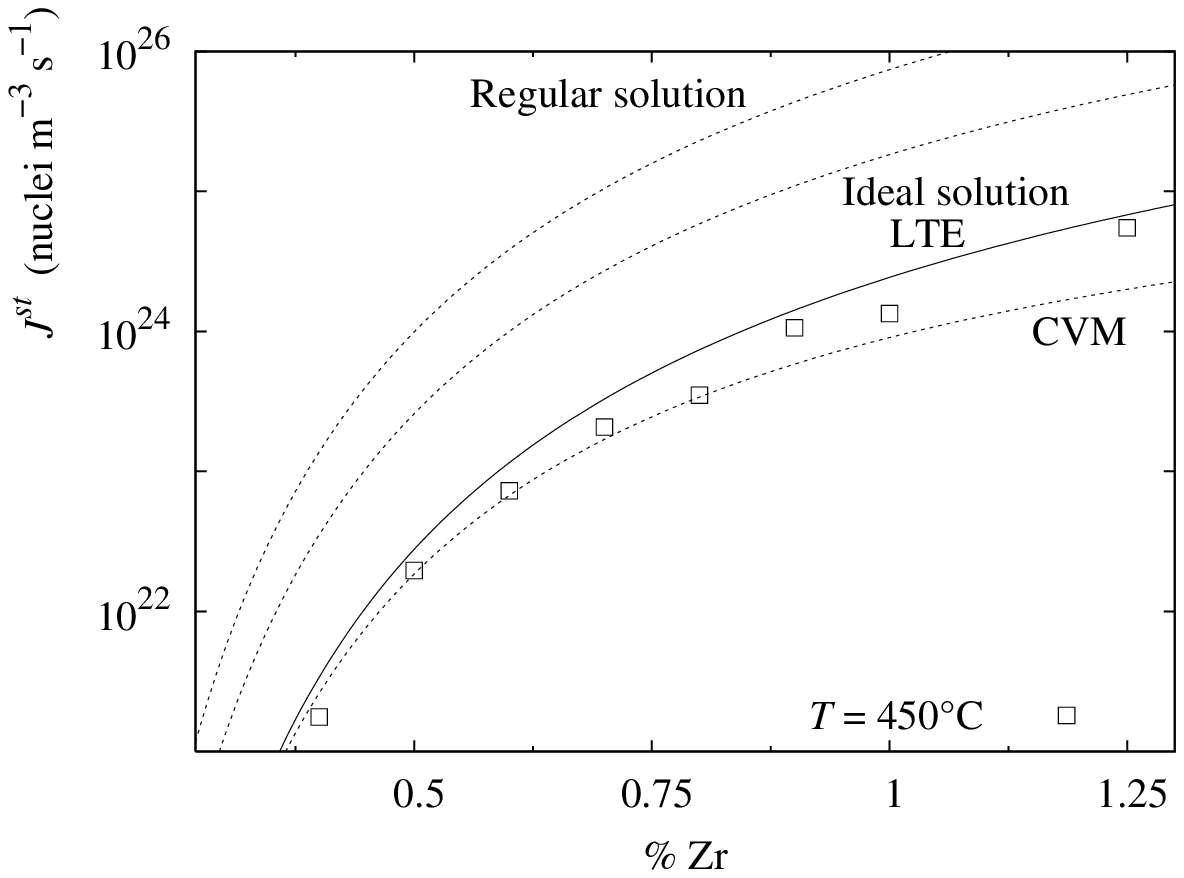}
		\hfill
		\includegraphics[width=0.49\linewidth]{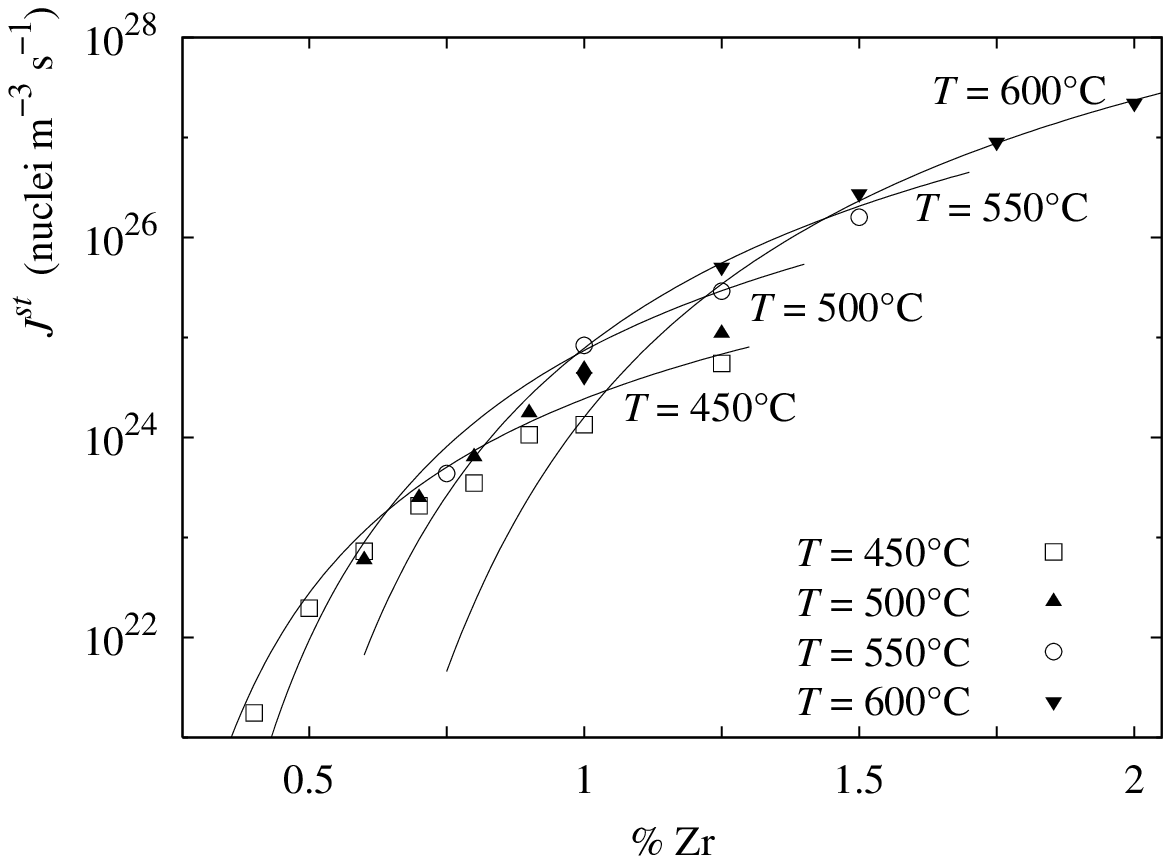}
	\end{center}
	\caption{Variation with the nominal concentration and the temperature of
	the steady-state nucleation rate $J^{st}$ for Al$_3$Zr
	precipitation. Symbols correspond to KMC simulations
	and lines to CNT.
	On the left, different thermodynamic approximations are used to evaluate
	$\Delta G^{nuc}$: the ideal and regular solid solution models,
	the cluster variation method (CVM) and the low temperature expansion
	(LTE) to second order (Eq.~\ref{eq:Gnuc_BT}).
	On the right, only LTE is used.}
	\label{fig:Jstat}
\end{figure}

Different thermodynamic approximations
can be used to obtain the nucleation free energy from the atomic model.
Usually, one uses simple mean-field approximation, like the ideal
or regular solid solution model which both rely on the Bragg-Williams
approximation. Nevertheless, we showed in Ref.~\citen{CLO04} 
that they lead to bad predictions of the CNT:
the cluster size distribution in the solid solution 
corresponding to Eq.~\ref{eq:distri_cnt} completely disagrees with that observed
during KMC simulations.
As the steady-state nucleation rate relies on this distribution,
we observe the same discrepancy for the prediction of $J^{st}$
(Fig.~\ref{fig:Jstat}).
So as to obtain a good agreement,
one has to take into account order effects corresponding to the strong
attraction existing between Al and Zr as well as Al and Sc atoms 
when they are first nearest neighbors and the strong repulsion
when they are second nearest neighbors
This can be done using the cluster variation method (CVM) \cite{KIK51,SAN78}
to calculate the nucleation free energy. 
The cluster size distribution given by Eq.~\ref{eq:distri_cnt}
then corresponds to the one observed in atomic simulations
and CNT theory well reproduces 
the nucleation rate (Ref.~\citen{CLO04} and Fig.~\ref{fig:Jstat}).
Nevertheless, one drawback of CVM is that it does not lead
to an analytical expression of the nucleation free energy.
It is then not really convenient to combine it with a theory
whose main contribution is to give an analytical expression 
of the nucleation rate.
Moreover, it would be interesting to understand thoroughly 
what is wrong with the ideal and the regular solid solution models
and in particular why the regular model leads to worse predictions
than the ideal one.
In this purpose, we use low temperature expansions\cite{DUCASTELLE}. 
This thermodynamic approximation leads to analytical expression 
of the nucleation free energy and takes correctly into account
the ordering tendency of the alloy.

A low temperature expansion consists in developing the partition 
function of the system around a reference state, keeping in the series
only the excited states of lowest energies. 
For an Al-Sc or an Al-Zr solid solution, this reference state corresponds
to pure Al and the two first excited states are the transformation
of an Al atom into an X solute atom and of a pair of Al second nearest neighbors
to a pair of X atoms.
When considering only the second order of the expansion, one can obtain 
analytical expressions of the different thermodynamic quantities in the canonic ensemble, 
thus depending on the nominal concentration $\xO$ of the solid solution.\footnote{A
detailed description of the calculations can be found in Ref.~\citen{CLO04T}.}
In particular, the nucleation free energy takes the form
\begin{equation}
	\label{eq:Gnuc_BT}
	\begin{split}
	\Delta G^{nuc}(\xO) =& kT \left[ \q{(\xO)}-\q{(\xeq)} \right] 
	+ 3kT \exp{\left(2\odeux/kT\right)}\left[ \q{(\xO)}^2-\q{(\xeq)^2} \right] \\
	&-\frac{1}{4} kT \left\{ \ln{\left[\q{(\xO)}\right]}-\ln{\left[\q{(\xeq)}\right]} \right\} ,
	\end{split}
\end{equation}
where we have defined the function
\begin{equation}
	\q{(x)}=\frac{2x}
	{1+\sqrt{1+24x\exp{\left(2\odeux/kT\right)}}}.
\end{equation}
Using this expression of the nucleation free energy with CNT
one obtains predictions of the nucleation rate as good
as with CVM (Fig.~\ref{fig:Jstat}) compared to the rate observed during
KMC simulations. 
In particular, the variations with the concentration of the solid solution
and with the annealing temperatures are well reproduced.

So as to understand why simple mean-field approximations lead to bad approximations
of the nucleation free energy, we develop the expression \ref{eq:Gnuc_BT} 
to first order in the concentrations $\xO$ and $\xeq$:
\begin{equation}
	\label{eq:Gnuc_BT_DL}
	\Delta G^{nuc}_{LTE}(\xO) \sim
	\frac{3}{4} kT \ln{\left( \frac{1-\xeq}{1-\xO} \right)}
	+ \frac{1}{4} kT \ln{\left( \frac{\xeq}{\xO} \right)} 
	+ \frac{1}{4}\left( 1+6\me^{2\odeux/kT} \right)\left( \xO - \xeq \right).
%	+ \frac{1}{4}\left( 1+6\exp{\left(\frac{2\odeux}{kT}\right)} \right)\left( \xO - \xeq \right).
\end{equation}
Doing the same development for the nucleation free energy calculated 
within the Bragg-Williams approximation (regular solid solution), we obtain
\begin{equation}
	\label{eq:Gnuc_BW_DL}
	\Delta G^{nuc}_{reg}(\xO) \sim
	\frac{3}{4} kT \ln{\left( \frac{1-\xeq}{1-\xO} \right)}
	+ \frac{1}{4} kT \ln{\left( \frac{\xeq}{\xO} \right)} 
	+ \left( 6 \oun + 3 \odeux \right)\left( \xO - \xeq \right).
\end{equation}
Comparing Eq.~\ref{eq:Gnuc_BT_DL} with Eq.~\ref{eq:Gnuc_BW_DL}, we see that 
these two thermodynamic approximations deviate from the ideal solid solution
model by a distinct linear term.
In the low temperature expansion (Eq.~\ref{eq:Gnuc_BT_DL}), the nucleation free energy is only
depending on the second nearest neighbor interaction and the coefficient
in front of the concentration is positive.
On the other hand, the Bragg-Williams approximation (Eq.~\ref{eq:Gnuc_BW_DL})
incorporates both first and second nearest neighbor interactions 
into a global parameter $\omega_{\mathrm{AlX}}=6\oun+3\odeux$.
This leads to a linear correction with a negative coefficient
as $\omega_{\mathrm{AlX}}<0$ for both binary alloys.
We thus demonstrate that the regular solid solution model leads to a wrong correction 
of the ideal model because it does not consider properly short range order.
In the case of a L1$_2$ ordered compound precipitating from a solid solution
lying on a fcc lattice, one cannot use such an approximation to calculate 
the nucleation free energy.
On the other hand, Eq.~\ref{eq:Gnuc_BT} is a good approximation
and can be used to calculate the nucleation free energy
even when the second nearest neighbor interaction $\odeux$ is not 
known. Indeed, this parameter can be deduced from the solubility limit
$\xeq$ by inverting Eq.~4 of Ref.~\cite{CLO04}, leading to the relation
\begin{equation}
	\odeux=-\frac{1}{6}kT\ln{\left( \xeq \right)}
	+ kT\left( {\xeq}^{2/3} + \frac{49}{6}{\xeq}^{4/3} \right).
	\label{eq:omega2_BT}
\end{equation}
This relation combined with Eq.~\ref{eq:Gnuc_BT} provides a powerful
way for calculating the nucleation free energy from the solid solubility.

\subsection{Ternary Al-Zr-Sc alloy}

We now explore how CNT can be extended to 
the Al-Zr-Sc alloy. 
Going from the binary to the ternary alloy, the main difficulty 
is that the precipitates have the composition Al$_3$Zr$_x$Sc$_{1-x}$,
where $x$ is not known a priori.
Therefore, the theory has to predict their composition.
As we showed for the two binary Al-Zr and Al-Sc alloys
that the capillary approximation is well suited to describe thermodynamics
of the clusters, even the smallest ones, We use the same approximation
for the ternary system. 
The formation free energy of a cluster Al$_3$Zr$_x$Sc$_{1-x}$
containing $\nX$ solute atoms can then be written
\begin{equation}
	\Delta G_{\nX}(x^0_{\mathrm{Zr}}, x^0_{\mathrm{Sc}}, x) 
	= 4\nX \Delta G^{nuc}(x^0_{\mathrm{Zr}}, x^0_{\mathrm{Sc}}, x)
	+ \left(36\pi\right)^{1/3}{\nX}^{2/3}a^2\bar{\sigma}(x).
	\label{eq:Gn_ternaire}
\end{equation}

The nucleation free energy is depending not only on the composition 
of the supersaturated solid solution but also on the composition of the nucleus: 
\begin{equation}
\begin{split}
	\Delta G^{nuc}(x^0_{\mathrm{Zr}}, x^0_{\mathrm{Sc}}, x) = &
	\frac{1}{4} \left\{ x \left[ 3\muAl^{p}(x) + \muZr^{p}(x) \right]
	+ (1-x) \left[ 3\muAl^{p}(x) + \muSc^{p}(x) \right] \right.\\
	& - x \left[ 3\muAl^{ss}(x^0_{\mathrm{Zr}}, x^0_{\mathrm{Sc}})
   	+ \muZr^{ss}(x^0_{\mathrm{Zr}}, x^0_{\mathrm{Sc}}) \right] \\
	&\left.- (1-x) \left[ 3\muAl^{ss}(x^0_{\mathrm{Zr}}, x^0_{\mathrm{Sc}})
   	+ \muSc^{ss}(x^0_{\mathrm{Zr}}, x^0_{\mathrm{Sc}}) \right]
	\right\}.
\end{split}
\label{eq:Gnuc_ternaire}
\end{equation}
$\muAl^{ss}$, $\muSc^{ss}$ and $\muZr^{ss}$ are the chemical potentials 
in the solid solution: we use a low temperature expansion to the second order
to calculate them.
$\muAl^p$, $\muSc^p$ and $\muZr^p$ are the chemical potentials in the precipitate.
The thermodynamic approach used to calculate them mixes low temperature expansion
so as to correctly describe the majority sublattice which only contains Al atoms 
as well as Bragg-Williams approximation because of the minority sublattice
which can equally be occupied by Zr or Sc atoms. This approach ensures that for $x=0$
or 1, Eq.~\ref{eq:Gnuc_ternaire} leads to the same value for the nucleation 
free energy as the calculation in the corresponding binary alloy (Eq.~\ref{eq:Gnuc_BT}).

As the two binary Al$_3$Zr and Al$_3$Sc compounds have a close interface 
free energy and as the interaction parameter $\omega^{(2)}_{\mathrm{ZrSc}}$ 
is low in magnitude, we assume a linear interpolation to calculate 
the interface free energy of the Al$_3$Zr$_x$Sc$_{1-x}$ compound,
\begin{equation}
	\bar{\sigma}(x) = x \bar{\sigma}(\mathrm{Al}_3\mathrm{Zr}) +
	(1-x)\bar{\sigma}(\mathrm{Al}_3\mathrm{Sc}).
	\label{eq:sigma_ternaire}
\end{equation}
CNT is not as sensitive as CD to this parameter because 
not all the thermodynamics is based on it.
Therefore, one does not need to consider the size dependence of the interface
free energy as we did for CD.

\begin{figure}[!bt]
	\begin{center}
		\includegraphics[width=0.6\textwidth]{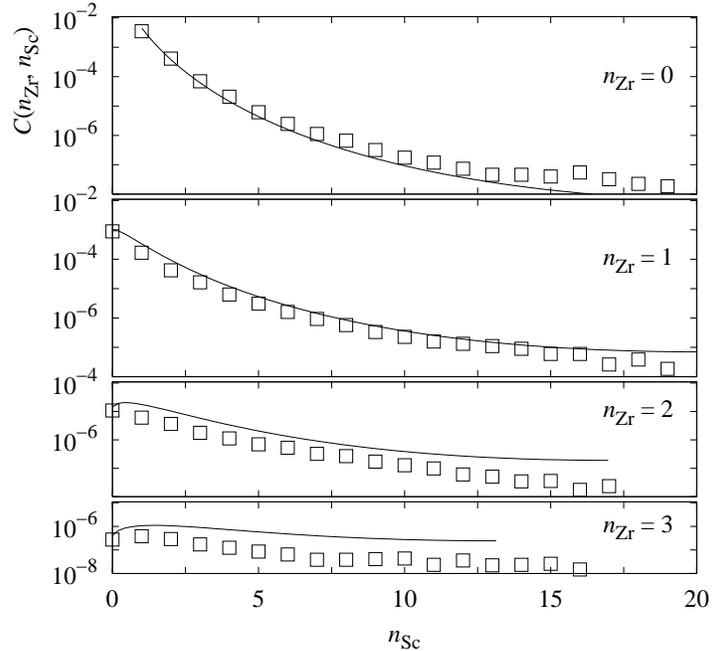}
	\end{center}
	\caption{Size distribution of Al$_3$Zr$_x$Sc$_{1-x}$ clusters 
	depending on their number $\nZr$ and $\nSc$ of atoms
	for a solid solution of concentration $x^0_{\mathrm{Zr}}=0.1$~ at.\%
	and $x^0_{\mathrm{Sc}}=0.5$~at.\% at $T=550^{\circ}$C.
	Symbols correspond to KMC simulations and lines to predictions 
	of CNT.}
	\label{fig:distri_ternaire}
\end{figure}

The cluster size distribution in the metastable solid solution 
still obeys Eq.~\ref{eq:distri_cnt} where now the cluster 
free energy is depending on the cluster composition and is given 
by Eq.~\ref{eq:Gn_ternaire}. 
Comparing this distribution with the one observed during
KMC simulations (Fig.~\ref{fig:distri_ternaire}), 
we obtain a satisfactory agreement showing that our thermodynamic
description of the cluster assembly is correct.

\begin{figure}[!bt]
	\begin{center}
		\includegraphics[width=0.5\textwidth]{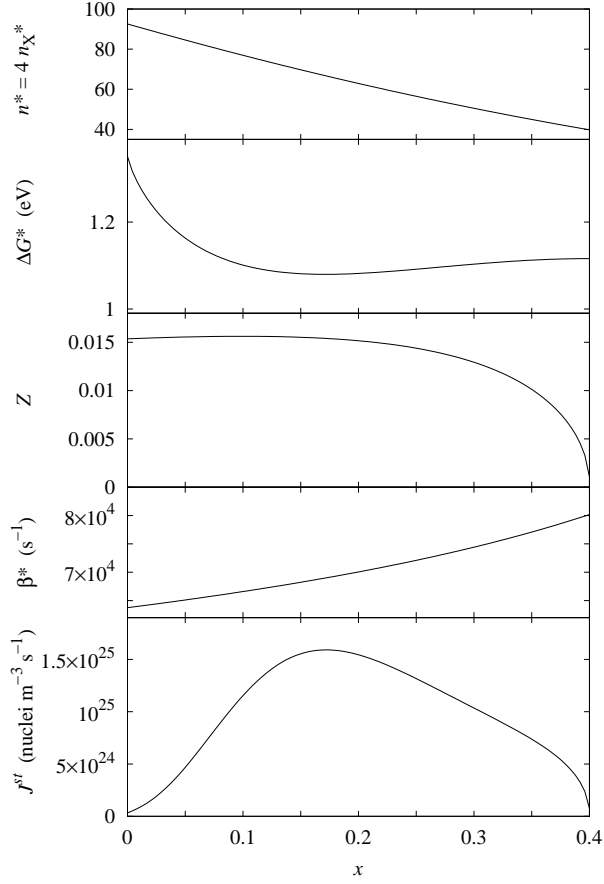}
	\end{center}
	\caption{Variation with the composition $x$ of the Al$_3$Zr$_x$Sc$_{1-x}$
	nucleus of the critical size $\nX^*$ (Eq~\ref{eq:nstar_ternaire}),
	the free energy $\Delta G^*$ of the critical nucleus,
	the Zeldovitch factor $Z$ (Eq.~\ref{eq:Zeldo_ternaire}),
	the condensation rate $\beta^*$ (Eq.~\ref{eq:beta_ternaire})
	and the nucleation rate $J^{st}$ (Eq.~\ref{eq:Jst_ternaire})
	for a solid solution of concentration $x^0_{\mathrm{Zr}}=0.1$~ at.\%
	and $x^0_{\mathrm{Sc}}=0.5$~at.\% at $T=550^{\circ}$C.}
	\label{fig:cinetique_ternaire}
\end{figure}

So as to define a classical nucleation rate we use the fact that 
Sc diffuses faster than Zr. We therefore assume that any critical 
nucleus is growing by absorbing a Sc atom and we neglect Zr absorption.
A critical size can then be defined for each set of clusters
containing the same number $\nZr$ of Zr atoms. This critical
size corresponds to the maximum of the cluster free energy
(Eq.~\ref{eq:Gn_ternaire}) with $\nZr$ being fixed and 
is thus given by the condition
\begin{equation}
	\left( \frac{\partial{\Delta G_{\nX}(x)}}
	{\partial{n_{\mathrm{Sc}}}}\right)_{n_{\mathrm{Zr}}}=0.
	\label{eq:Gn_deriv_ternaire}
\end{equation}
This leads to a critical size depending on the composition $x$
of the cluster,
\begin{equation}
	n^*(x) = 4\nX^*(x) = \frac{16\pi}{3} \left( \frac{(2+x)\
	a^2\bar{\sigma}(\mathrm{Al}_3\mathrm{Sc}) - x\
	a^2\bar{\sigma}(\mathrm{Al}_3\mathrm{Zr})} {\muSc^{ss}+3\
	\muAl^{ss} - \muSc^p(x)-3\ \muAl^p(x)} \right)^3.
	\label{eq:nstar_ternaire}
\end{equation}

We can define now a nucleation rate $J^{st}(x)$ 
depending on the composition of the nucleus. 
As in the binary alloy, this rate takes the form
\begin{equation}
	J^{st}(x) = N_s \ Z(x) \ \beta^*(x) \ \exp{\left(-\Delta G^*(x)/kT \right)}.
	\label{eq:Jst_ternaire}
\end{equation}
The Zeldovitch factor $Z(x)$ is defined along the destabilization direction 
of the nucleus and is thus obtained by considering the second derivative 
of the nucleus free energy with respect to $\nSc$, keeping $\nZr$ constant:
\begin{equation}
	Z(x) = 4\sqrt{ -\frac{1}{32\pi kT}
	\left(\frac{\partial^2{\Delta G_n}}
	{\partial{n_{\mathrm{Sc}}}^2}
	\right)_{n_{\mathrm{Zr}}} }.
	\label{eq:Zeldo_ternaire}
\end{equation}
$\Delta G^*(x)$ is the cluster free energy (Eq.~\ref{eq:Gn_ternaire})
of a cluster having the critical size $\nX^*(x)$.
As for the condensation rate, it is obtained by considering the absorption 
of Sc atoms by the critical nucleus, leading to
\begin{equation}
	\beta^*(x) = 8 \pi \left( \frac{3\nX^*(x)}{2\pi} \right)^{1/3}
	\frac{D_{\mathrm{Sc}}}{a^2}\frac{x^0_{\mathrm{Sc}}}{1-x}.
	\label{eq:beta_ternaire}
\end{equation}
Writing the nucleation rate in this way, we ensure that when $x=0$ 
we obtain the same expression as that given by CNT
in the Al-Sc alloy (Eq.~\ref{eq:Jst_cnt}).

\begin{figure}[!bt]
	\begin{center}
		\includegraphics[width=0.5\linewidth]{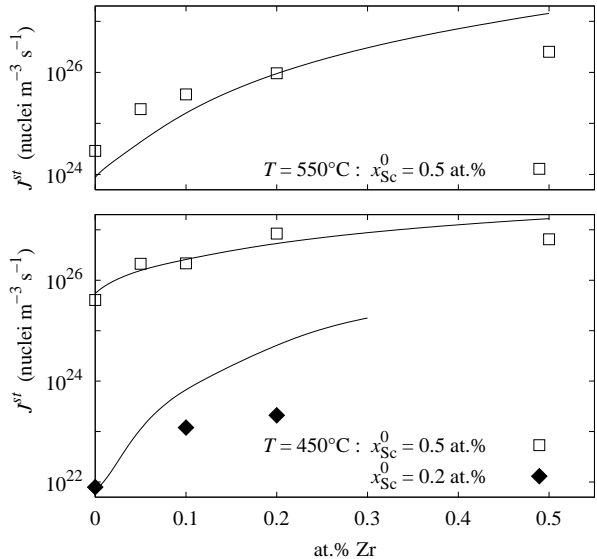}
	\end{center}
	\caption{Variation of the nucleation rate $J^{st}$ (Maximum of $J^{st}(x)$ 
	given by Eq.~\ref{eq:Jst_ternaire})
	with the zirconium concentration for different scandium concentrations of the 
	solid solution and different annealing temperatures.}
	\label{fig:Jst_ternaire}
\end{figure}

The variations of the nucleation rate $J^{st}(x)$ with the composition
$x$ of the nucleus are shown in Fig.~\ref{fig:cinetique_ternaire}
for a chosen supersaturated solid solution.
So as to compare results of this nucleation model with KMC simulations, 
we define the nucleation rate $J^{st}$ as the maximum of the function 
$J^{st}(x)$ given by Eq.~\ref{eq:Jst_ternaire}. 
The critical size used to extract the nucleation rate from KMC simulations
is the one corresponding to this maximum.
When the solid solution is highly supersaturated in Sc ($x^0_{\mathrm{Sc}}=0.5$~at.\%
at $T=450$ of $550^{\circ}$), our mesoscopic models reproduces the variation 
of the nucleation rate with the zirconium concentration of the solid solution 
(Fig.~\ref{fig:Jst_ternaire}).
The mesoscopic evaluation of $J^{st}$ worsens a little bit when the 
Zr concentration becomes comparable to the Sc one and when the Sc supersaturation
becomes lower. 
Nevertheless, the key point is that this extension of CNT manages to predict 
the increase of the nucleation rate with a Zr addition to an Al-Sc alloy,
this increase being in reasonable agreement with atomic simulations.
More advanced models (e.g. the link flux analysis and other approximations 
as described in Ref. \cite{REI50,HIR74,WU93}) are worth to be tried.

\section{Conclusions}

This study illustrates how a quantitative multiscale modeling 
of the precipitation kinetics can be performed. 
Using a very limited number of input data, 
we built  for the Al-Zr-Sc alloy an atomic model 
from which mesoscopic quantities like the interface 
free energy or the nucleation free energy could be deduced.
For the two binary Al-Zr and Al-Sc alloys, it was shown
that a good agreement can be obtained between the KMC simulations,
different mesoscopic models (CD and CNT) and experimental data. 
So as to obtain good predictions with CNT,
one has to take into account the ordering tendency 
of the system.
In this purpose, using low temperature 
expansions, we propose an analytical expression 
of the nucleation free energy.
For the ternary Al-Zr-Sc alloy, we showed that the precipitate
inhomogeneity arises from the difference between the two solute diffusion coefficients.
Although the precipitate composition is not known a priori, 
CNT could be extended to this ternary alloy thanks to reasonable 
assumptions on the kinetics. 
This mesoscopic model reproduces the increase of the nucleation 
rate associated with a Zr addition to an Al-Sc alloy.

\section{Acknowledgments}
The authors are grateful to Prof. A.J. Ardell, Prof. D.N. Seidman
and Dr. J. R{\o}yset for providing experimental data.
They wish to thank Dr. B. Legrand and Dr. F. Soisson 
for very fruitful discussions.
Part of this work has been funded by the joint research program
``Precipitation'' between Alcan, Arcelor, CNRS, and CEA. 
\bibliographystyle{tms}
\bibliography{clouet2005}

\begin{thebibliography}{10}
\newcommand{\enquote}[1]{``#1''}

\bibitem{YEL85}
V.~I. Yelagin, V.~V. Zakharov, S.~G. Pavlenko, and T.~D. Rostova.
\newblock \enquote{Influence of Zirconium Additions on Ageing of {A}l-{S}c
  Alloys.}
\newblock \emph{Phys. Met. Metall.}, 60 (1985), 88--92

\bibitem{DAV96}
V.~G. Davydov, V.~I. Yelagin, V.~V. Zakharov, and T.~D. Rostova.
\newblock \enquote{Alloying Aluminum Alloys with Scandium and Zirconium
  Additives.}
\newblock \emph{Metal Science and Heat Treatment}, 38 (1996), 347--352

\bibitem{TOR98}
L.~S. Toropova, D.~G. Eskin, M.~L. Kharaterova, and T.~V. Bobatkina.
\newblock \emph{Advanced Aluminum Alloys Containing Scandium - Structure and
  Properties} (Amsterdam: Gordon and Breach Sciences, 1998)

\bibitem{FUL03}
C.~B. Fuller, D.~N. Seidman, and D.~C. Dunand.
\newblock \enquote{Mechanical Properties of {A}l({S}c,{Z}r) Alloys at Ambient
  and Elevated Temperatures.}
\newblock \emph{Acta Mater.}, 51 (2003), 4803--4814

\bibitem{RID04}
Y.~W. Riddle and T.~H. Sanders.
\newblock \enquote{A Study of Coarsening, Recrystallization, and Morphology of
  Microstructure in {A}l-{S}c-({Z}r)-({M}g) Alloys.}
\newblock \emph{Metall. Mater. Trans. A}, 35 (2004), 341--350

\bibitem{ROY05b}
J.~R{\o}yset and N.~Ryum.
\newblock \enquote{Scandium in Aluminium Alloys.}
\newblock \emph{Int. Mater. Rev.}, 50 (2005), 19--44

\bibitem{PEARSON}
P.~Villars and L.~D. Calvert.
\newblock \emph{Pearson's Handbook of Crystallographic Data for Intermetallic
  Phases} (Oh: American Society for Metals, 1985)

\bibitem{HAR02}
Y.~Harada and D.~C. Dunand.
\newblock \enquote{Microstructure of {A}l$_3${S}c with Ternary Transition-Metal
  Additions.}
\newblock \emph{Mater. Sci. Eng.}, A329--331 (2002), 686--695

\bibitem{ROB01}
J.~D. Robson and P.~B. Prangnell.
\newblock \enquote{Dispersoid Precipitation and Process Modelling in Zirconium
  Containing Commercial Aluminium Alloys.}
\newblock \emph{Acta Mater.}, 49 (2001), 599--613

\bibitem{RYU69}
N.~Ryum.
\newblock \enquote{Precipitation and Recrystallization in an
  {A}l-0.5~wt.\%~{Z}r Alloy.}
\newblock \emph{Acta Metall.}, 17 (1969), 269--278

\bibitem{NES72}
E.~Nes.
\newblock \enquote{Precipitation of the Metastable Cubic {A}l$_3${Z}r-Phase in
  Subperitectic {A}l-{Z}r Alloys.}
\newblock \emph{Acta Metall.}, 20 (1972), 499--506

\bibitem{CLO04}
E.~Clouet, M.~Nastar, and C.~Sigli.
\newblock \enquote{Nucleation of {Al$_3$Zr} and {Al$_3$Sc} in Aluminum Alloys:
  from Kinetic {M}onte {C}arlo Simulations to Classical Theory.}
\newblock \emph{Phys. Rev. B}, 69 (2004), 064109

\bibitem{CLO02}
E.~Clouet, J.~M. Sanchez, and C.~Sigli.
\newblock \enquote{First-Principles Study of the Solubility of {Z}r in {A}l.}
\newblock \emph{Phys. Rev. B}, 65 (2002), 094105

\bibitem{MET93}
M.~Methfessel and M.~van Schilfgaarde.
\newblock \enquote{Derivation of Force Theorem in Density-Functional Theory:
  Application to the Full-Potential {LMTO} Method.}
\newblock \emph{Phys. Rev. B}, 48 (1993), 4937--4940

\bibitem{PER96}
J.~P. Perdew, K.~Burke, and M.~Ernzerhof.
\newblock \enquote{Generalized Gradient Approximation Made Simple.}
\newblock \emph{Phys. Rev. Lett.}, 77 (1996), 3865--3868

\bibitem{CON83}
J.~W. Connolly and A.~R. Williams.
\newblock \enquote{Density-Functional Theory Applied to Phase Transformations
  in Transition-Metal Alloys.}
\newblock \emph{Phys. Rev. B}, 27 (1983), 5169--5172

\bibitem{TOR90}
L.~S. Toropova, A.~N. Kamardinkin, V.~V. Kindzhibalo, and A.~T. Tyvanchuk.
\newblock \enquote{Investigation of Alloys of the {A}l-{S}c-{Z}r System in the
  Aluminium-Rich Range.}
\newblock \emph{Phys. Met. Metall.}, 70 (1990), 106--110

\bibitem{TOL05}
A.~Tolley, V.~Radmilovic, and U.~Dahmen.
\newblock \enquote{Segregation in {A}l$_3$({S}c,{Z}r) Precipitates in
  {A}l-{S}c-{Z}r Alloys.}
\newblock \emph{Scripta Mater.}, 52 (2005), 621--625

\bibitem{RAD05}
V.~Radmilovic, A.~Tolley, and U.~Dahmen.
\newblock \enquote{Precipitate Evolution and Coarsening Kinetics in Al-Sc-Zr
  Alloys.}
\newblock In \enquote{Solid-Solid Phase Transformations in Inorganic
  Materials,}  (TMS, 2005)

\bibitem{FOR04}
B.~Forbord, W.~Lefebvre, F.~Danoix, H.~Hallem, and K.~Marthinsen.
\newblock \enquote{Three Dimensional Atom Probe Investigation on the Formation
  of {A}l$_3$({S}c,{Z}r)-Dispersoids in Aluminium Alloys.}
\newblock \emph{Scripta Mater.}, 51 (2004), 333--337

\bibitem{FUL05a}
C.~B. Fuller, J.~L. Murray, and D.~N. Seidman.
\newblock \enquote{Temporal Evolution of the Nanostructure of Al(Sc,Zr) Alloys:
  Part I-Chemical Compositions of Al$_3$(Sc$_{1-x}$Zr$_x$) Precipitates.}
\newblock \emph{Acta Mater.}, submitted (2005)

\bibitem{SOI00}
F.~Soisson and G.~Martin.
\newblock \enquote{{M}onte-{C}arlo Simulations of the Decomposition of
  Metastable Solid Solutions: Transient and Steady-State Nucleation Kinetics.}
\newblock \emph{Phys. Rev. B}, 62 (2000), 203--214

\bibitem{CLO05}
E.~Clouet, A.~Barbu, L.~Laé, and G.~Martin.
\newblock \enquote{Precipitation Kinetics of {A}l$_3${Z}r and {A}l$_3${S}c in
  Aluminum Alloys Modeled with Cluster Dynamics.}
\newblock \emph{Acta Mater.}, 53 (2005), 2313--2325

\bibitem{GOL95}
S.~I. Golubov, Y.~N. Osetsky, A.~Serra, and A.~V. Barashev.
\newblock \enquote{The Evolution of Copper Precipitates in Binary {F}e-{C}u
  Alloys during Ageing and Irradiation.}
\newblock \emph{J. Nucl. Mater.}, 226 (1995), 252--255

\bibitem{MAT97}
M.~H. Mathon, A.~Barbu, F.~Dunstetter, F.~Maury, N.~Lorenzelli, and C.~H. {de
  Novion}.
\newblock \enquote{Experimental Study and Modelling of Copper Precipitation
  Under Irradiation in Dilute {F}e{C}u Alloys.}
\newblock \emph{J. Nucl. Mater.}, 245 (1997), 224--237

\bibitem{BAR04}
A.~V. Barashev, S.~I. Golubov, D.~J. Bacon, P.~E.~J. Flewitt, and T.~A. Lewis.
\newblock \enquote{Copper Precipitation in {F}e-{C}u Alloys under Electron and
  Neutron Irradiation.}
\newblock \emph{Acta Mater.}, 52 (2004), 877--886

\bibitem{MAR05}
G.~Martin.
\newblock \enquote{Reconciling the Classical Theory of Nucleation and Atomic
  Scale-Observations and Modeling.}
\newblock In \enquote{Solid-Solid Phase Transformations in Inorganic
  Materials,}  (TMS, 2005)

\bibitem{WAI58}
T.~R. Waite.
\newblock \enquote{General Theory of Bimolecular Reaction Rates in Solids and
  Liquids.}
\newblock \emph{J. Chem. Phys.}, 28 (1958), 103--106

\bibitem{MAR78}
G.~Martin.
\newblock \enquote{The Theories of Unmixing Kinetics of Solid Solutions.}
\newblock In \enquote{Solid State Phase Transformation in Metals and Alloys,}
  (Orsay, France: Les \'Editions de Physique, 1978), 337--406

\bibitem{MAR01}
E.~A. Marquis and D.~N. Seidman.
\newblock \enquote{Nanoscale Structural Evolution of {A}l$_3${S}c Precipitates
  in {A}l-{S}c Alloys.}
\newblock \emph{Acta Mater.}, 49 (2001), 1909--1919

\bibitem{PORTER}
D.~A. Porter and K.~E. Easterling.
\newblock \emph{Phase Transformations in Metals and Alloys} (London: Chapman \&
  Hall, 1992)

\bibitem{CHRISTIAN}
J.~W. Christian.
\newblock \emph{The Theory of Transformations in Metals and Alloys - Part~{I}:
  Equilibrium and General Kinetic Theory} (Oxford: Pergamon Press, 1975)

\bibitem{PER84}
A.~Perini, G.~Jacucci, and G.~Martin.
\newblock \enquote{Cluster Free Energy in the Simple-Cubic {I}sing Model.}
\newblock \emph{Phys. Rev. B}, 29 (1984), 2689--2697

\bibitem{NOV01}
G.~M. Novotny and A.~J. Ardell.
\newblock \enquote{Precipitation of {A}l$_3${S}c in Binary {A}l-{S}c Alloys.}
\newblock \emph{Mater. Sci. Eng. A}, 318 (2001), 144--154

\bibitem{MAR02}
E.~A. Marquis, D.~N. Seidman, and D.~C. Dunand.
\newblock \enquote{Creep of Precipitation-Strengthened {A}l({S}c) Alloys.}
\newblock In R.~S. Mishra, J.~C. Earthman, and S.~V. Raj, eds., \enquote{Creep
  Deformation: Fundamentals and Applications,}  (TMS, 2002), 299

\bibitem{MAR02T}
E.~A. Marquis.
\newblock \emph{Microstructural Evolution and Strengthening Mechanism in
  {A}l-{S}c and {A}l-{M}g-{S}c Alloys}.
\newblock Ph.D. thesis, Northwestern University, Evanston, Illinois (2002)

\bibitem{ZAK97}
V.~V. Zakharov.
\newblock \enquote{Stability of the Solid Solution of Scandium in Aluminum.}
\newblock \emph{Metal Science and Heat Treatment}, 39 (1997), 61--66

\bibitem{ROY05c}
J.~R{\o}yset and N.~Ryum.
\newblock \enquote{Kinetics and Mechanisms of Precipitation in an {A}l-0.2
  wt.\% {S}c Alloy.}
\newblock \emph{Mater. Sci. Eng. A}, 396 (2005), 409--422

\bibitem{LUI80}
N.~Luiggi, J.~P. Simon, and P.~Guyot.
\newblock \enquote{Residual Resistivity of Clusters in Solid Solutions.}
\newblock \emph{J. Phys. F}, 10 (1980), 865--872

\bibitem{ROS87}
P.~L. Rossiter.
\newblock \emph{The Electrical Resistivity of Metals and Alloys} (Cambridge:
  Cambridge Universtiy Press, 1987)

\bibitem{FUJ79}
S.~I. Fujikawa, M.~Sugaya, H.~Takei, and K.~I. Hirano.
\newblock \enquote{Solid Solubility and Residual Resistivity of Scandium in
  Aluminium.}
\newblock \emph{J. Less-Common Met.}, 63 (1979), 87--97

\bibitem{JO93}
H.-H. Jo and S.-I. Fujikawa.
\newblock \enquote{Kinetics of Precipitation in {A}l-{S}c Alloys and Low
  Temperature Solid Solubility of Scandium in Aluminium studied by Electrical
  Resistivity Measurements.}
\newblock \emph{Mater. Sci. Eng. A}, 171 (1993), 151--161

\bibitem{FLYNN}
C.~P. Flynn.
\newblock \emph{Point Defects and Diffusion} (Oxford: Clarendon Press, 1972)

\bibitem{LAE05}
L.~Laé and P.~Guyot.
\newblock \enquote{Cluster Dynamics Modeling of Precipitation Kinetics: Recent
  Developments.}
\newblock In \enquote{Solid-Solid Phase Transformations in Inorganic
  Materials,}  (TMS, 2005)

\bibitem{WU97}
D.~T. Wu.
\newblock \enquote{Nucleation Theory.}
\newblock \emph{Solid State Physics}, 50 (1997), 37--187

\bibitem{WAG91}
R.~Wagner and R.~Kampmann.
\newblock \enquote{Homogeneous Second Phase Precipitation.}
\newblock In R.~W. Cahn, P.~Haasen, and E.~J. Kramer, eds., \enquote{Materials
  Science and Technology, a Comprehensive Treatment,}  (Weinheim: VCH, 1991),
  volume~5, chapter~4. 213--303

\bibitem{RAM99}
R.~A. Ramos, P.~A. Rikvold, and M.~A. Novotny.
\newblock \enquote{Test of the {K}olmogorov-{J}ohnson-{M}ehl-{A}vrami Picture
  of Metastable Decay in a Model with Microscopic Dynamics.}
\newblock \emph{Phys. Rev. B}, 59 (1999), 9053--9069

\bibitem{SHN99}
V.~A. Shneidman, K.~A. Jackson, and K.~M. Beatty.
\newblock \enquote{Nucleation and Growth of a Stable Phase in an {I}sing-type
  System.}
\newblock \emph{Phys. Rev. B}, 59 (1999), 3579--3589

\bibitem{NOV00}
M.~A. Novotny, P.~A. Rikvold, M.~Kolesik, D.~M. Townsley, and R.~A. Ramos.
\newblock \enquote{Simulations of Metastable Decay in Two- and
  Three-Dimensional Models with Microscopic Dynamics.}
\newblock \emph{J. Non-Cryst. Solids}, 274 (2000), 356--363

\bibitem{BER04}
F.~Berthier, B.~Legrand, J.~Creuze, and R.~T\'etot.
\newblock \enquote{Atomistic Investigation of the
  {K}olmogorov-{J}ohnson-{M}ehl-{A}vrami Law in Electrodeposition Process.}
\newblock \emph{J. Electroanal. Chem.}, 561 (2004), 37

\bibitem{BER04b}
F.~Berthier, B.~Legrand, J.~Creuze, and R.~T\'etot.
\newblock \enquote{{A}g/{C}u (001) Electrodeposition: Beyond the Classical
  Nucleation Theory.}
\newblock \emph{J. Electroanal. Chem.}, 562 (2004), 127

\bibitem{KIK51}
R.~Kikuchi.
\newblock \enquote{A Theory of Cooperative Phenomena.}
\newblock \emph{Phys. Rev.}, 81 (1951), 988--1003

\bibitem{SAN78}
J.~M. Sanchez and D.~de~Fontaine.
\newblock \enquote{The fcc {I}sing Model in the Cluster Variation
  Approximation.}
\newblock \emph{Phys. Rev. B}, 17 (1978), 2926--2936

\bibitem{DUCASTELLE}
F.~Ducastelle.
\newblock \emph{Order and Phase Stability in Alloys} (North-Holland, Amsterdam,
  1991)

\bibitem{CLO04T}
E.~Clouet.
\newblock \emph{Séparation de Phase dans les Alliages {A}l-{Z}r-{S}c: du Saut
  des Atomes à la Croissance de Précipités Ordonnés}.
\newblock Ph.D. thesis, \'Ecole Centrale Paris (2004).
\newblock Http://tel.ccsd.cnrs.fr/documents/archives0/00/00/59/74

\bibitem{REI50}
H.~Reiss.
\newblock \enquote{The Kinetics of Phase Transitions in Binary Systems.}
\newblock \emph{J. Chem. Phys.}, 18 (1950), 840--848

\bibitem{HIR74}
J.~O. Hirschfelder.
\newblock \enquote{Kinetics of Homogeneous Nucleation on Many-Component
  Systems.}
\newblock \emph{J. Chem. Phys.}, 61 (1974), 2690--2694

\bibitem{WU93}
D.~T. Wu.
\newblock \enquote{General Approach to Barrier Crossing in Multicomponent
  Nucleation.}
\newblock \emph{J. Chem. Phys.}, 99 (1993), 1990--2000

\end{thebibliography}

\end{document}